\documentclass[lineno]{jfm}
\usepackage{graphicx}
\usepackage{newtxtext}
\usepackage{newtxmath}
\usepackage{natbib}
\usepackage{hyperref}
\usepackage{soul}
\hypersetup{
    colorlinks = true,
    urlcolor   = blue,
    citecolor  = blue,
}

\newcommand{\RomanNumeralCaps}[1]
\linenumbers
\usepackage{tikz}
\newcommand*\circled[1]{\tikz[baseline=(char.base)]{
            \node[shape=circle,draw,inner sep=0.5pt] (char) {#1};}}

\shorttitle{Flow-induced oscillations of pitching swept wings}
\shortauthor{Y. Zhu and K. Breuer}

\title{\vspace{-3em}Flow-induced oscillations of pitching swept wings: Stability boundary, vortex dynamics and force partitioning}

\author{Yuanhang Zhu\aff{1}
  \corresp{\email{yuanhang\_zhu@brown.edu}},
  \and Kenneth Breuer\aff{1}
 }

\affiliation{\aff{1}
Center for Fluid Mechanics, School of Engineering, Brown University, Providence, RI 02912, USA}

\begin{document}
\maketitle

\begin{abstract}
We experimentally study the aeroelastic instability boundaries and three-dimensional vortex dynamics of pitching swept wings, with the sweep angle ranging from 0 to 25 degrees. The structural dynamics of the wings are simulated using a cyber-physical control system. With a constant flow speed, a prescribed high inertia and a small structural damping, we show that the system undergoes a subcritical Hopf bifurcation to large-amplitude limit-cycle oscillations (LCOs) for all the sweep angles. The onset of LCOs depends largely on the static characteristics of the wing. The saddle-node point is found to change non-monotonically with the sweep angle, which we attribute to the non-monotonic power transfer between the ambient fluid and the elastic mount. An optimal sweep angle is observed to enhance the power extraction performance and thus promote LCOs and destabilize the aeroelastic system. The frequency response of the system reveals a structural-hydrodynamic oscillation mode for wings with relatively high sweep angles. Force, moment, and three-dimensional flow structures measured using multi-layer stereoscopic particle image velocimetry are analyzed to explain the differences in power extraction for different swept wings. Finally, we employ a physics-based Force and Moment Partitioning Method (FMPM) to quantitatively correlate the three-dimensional vortex dynamics with the resultant unsteady aerodynamic moment.

\end{abstract}


\section{Introduction}\label{sec.intro}

The fluid-structure interaction (FSI) of elastically mounted pitching wings can lead to large-amplitude flow-induced oscillations under certain operating conditions. In extreme cases, these flow-induced oscillations may affect structural integrity and even cause catastrophic aeroelastic failures \citep{dowell1989modern}. On the other hand, however, hydro-kinetic energy can be harnessed from these oscillations, providing an alternative solution for next-generation renewable energy devices \citep{xiao2014review,young2014review,boudreau2018experimental,su2019resonant}. Moreover, the aero-/hydro-elastic interactions of passively pitching wings/fins have important connections with animal flight \citep{wang2005dissecting,bergou2007passive,beatus2015wing,wu2019scaling} and swimming \citep{long1996importance,quinn2021tunable}, and understanding these interactions may further aid the design and development of flapping-wing micro air vehicles (MAVs) \citep{shyy2010recent,jafferis2019untethered} and oscillating-foil autonomous underwater vehicles (AUVs) \citep{zhong2021tunable,tong2022design}.

Flow-induced oscillations of pitching wings originate from the two-way coupling between the structural dynamics of the elastic mount and the fluid force exerted on the wing. While the dynamics of the elastic mount can be approximated by a simple spring-mass-damper model, the fluid forcing term is usually found to be highly nonlinear due to the formation, growth, and shedding of a strong leading-edge vortex (LEV) \citep{mccroskey1982unsteady,dimitriadis2009bifurcation,mulleners2012onset,eldredge2019leading}. \citet{onoue2015large} and \citet{onoue2016vortex} experimentally studied the flow-induced oscillations of a pitching plate whose structural stiffness, damping and inertia were defined using a cyber-physical system (\S \ref{sec.CPS}, see also \citet{hover1997vortex,mackowski2011developing,zhu2020nonlinear}) and, using this approach, identified a subcritical bifurcation to aeroelastic instability. The temporal evolution of the LEV associated with the aeroelastic oscillations was characterized using particle image velocimetry (PIV), and the unsteady flow structures were correlated with the unsteady aerodynamic moments using a potential flow model. \citet{menon2019flow} numerically studied a similar problem, simulating an elastically mounted two-dimensional NACA-0015 airfoil at a Reynolds number of 1000. An energy approach, which bridges prescribed sinusoidal oscillations and passive flow-induced oscillations, was employed to characterize the dynamics of the aeroelastic system. The energy approach maps out the energy transfer between the ambient flow and the elastic mount over a range of prescribed pitching amplitudes and frequencies and unveils the system stability based on the sign of the energy gradient. 

More recently, \citet{zhu2020nonlinear} characterized the effect of wing inertia on the flow-induced oscillations of pitching wings and the corresponding LEV dynamics. Two distinct oscillation modes were reported: (i) a structural mode, which occurred via a subcritical bifurcation and was associated with a high-inertia wing, and (ii) a hydrodynamic mode, which occurred via a supercritical bifurcation and was associated with a low-inertia wing. The wing was found to shed one strong LEV during each half-pitching cycle for the hydrodynamic mode, whereas a weak secondary LEV was also shed in the high-inertial structural mode.

These previous studies have collectively demonstrated that LEV dynamics play an important role in shaping flow-induced oscillations and thus regulate the stability characteristics of passively pitching wings. However, these studies have only focused on studying the structural and flow dynamics of two-dimensional wings or airfoils. The extent to which these important findings for two-dimensional wings hold in three dimensions remains unclear.

Swept wings are commonly seen for flapping-wing fliers and swimmers in nature \citep{ellington1996leading,lentink2007swifts,borazjani2013fish,bottom2016hydrodynamics,zurman2021fin}, as well as on many engineered fixed-wing flying vehicles. It is argued that wing sweep can enhance lift generation for flapping wings because it stabilizes the LEV by maintaining its size through spanwise vorticity transport -- a mechanism similar to the lift enhancement mechanism of delta wings \citep{polhamus1971predictions}. \citet{chiereghin2020three} found significant spanwise flow for a high-aspect ratio plunging swept wing at a sweep angle of 40 degrees. In another study, for the same sweep angle, attached LEVs and vortex breakdown were observed just like those on delta wings \citep{gursul2019plunging}. Recent works have shown that the effect of wing sweep on LEV dynamics depends strongly on wing kinematics. \citet{beem2012stabilization} showed experimentally that for a plunging swept wing, the strong spanwise flow induced by the wing sweep is not sufficient for LEV stabilization. \citet{wong2013investigation} reinforced this argument by comparing the LEV stability of plunging and flapping swept wings and showed that two-dimensional (i.e. uniform without any velocity gradient) spanwise flow alone cannot stabilize LEVs -- there must be spanwise gradients in vorticity or spanwise flow so that vorticity can be convected or stretched. \citet{wong2015determining} demonstrated both theoretically and experimentally that the wing sweep improves relative LEV stability of flapping swept wings by enhancing the spanwise vorticity convection and stretching so as to keep the LEV size below a critical shedding threshold \citep{rival2014characteristic}. \citet{onoue2017scaling} experimentally studied elastically mounted pitching unswept and swept wings and proposed a universal scaling for the LEV formation time and circulation, which incorporated the effects of the pitching frequency, the pivot location and the sweep angle. The vortex circulation was demonstrated to be independent of the three-dimensional vortex dynamics. In addition, they concluded that the stability of LEV can be improved by moderating the LEV circulation through vorticity annihilation, which is largely governed by the shape of the leading-edge sweep, agreeing with the results of \citet{wojcik2014vorticity}. More recently, \citet{visbal2019effect} numerically studied the effect of wing sweep on the dynamic stall of pitching three-dimensional wings and reported that the wing sweep can modify the LEV structures and change the net aerodynamic damping of the wing. The effect of wing sweep on the LEV dynamics and stability, as one can imagine, will further affect the unsteady aerodynamic forces and thereby the aeroelastic response of pitching swept wings.

Another important flow feature associated with unsteady three-dimensional wings is the behavior of the tip vortex (TV). Although the tip vortex usually grows distinctly from the leading-edge vortex for rectangular platforms \citep{taira2009three,kim2010experimental,hartloper2013competition}, studies have suggested that the TV is able to anchor the LEV in the vicinity of the wing tip, which delays LEV shedding \citep{birch2001spanwise,hartloper2013competition}. Moreover, the tip vortex has also been shown to affect the unsteady wake dynamics of both unswept and swept wings \citep{taira2009three,zhang2020laminar,zhang2020formation,ribeiro2022wing,son2022leading,son2022dynamics}. However, it remains elusive how the interactions between LEVs and TVs change with the wing sweep, and more importantly, how this change will in turn affect the response of aeroelastic systems.

To dissect the effects of complex vortex dynamics associated with unsteady wings/airfoils, a physics-based Force and Moment Partitioning Method (FMPM) has been proposed \citep{quartapelle1983force,zhang2015centripetal,moriche2017aerodynamic,menon2021initiation,menon2021quantitative,menon2021significance} (also known as the vortex force/moment map method \citep{li2018vortex,li2020vortex}). The method has attracted attention recently due to its high versatility for analyzing a variety type of vortex-dominated flows. Under this framework, the Navier-Stokes equation is projected onto the gradient of an influence potential to separate the force contributions from the added-mass, vorticity-induced, and viscous terms. It is particularly useful for analyzing vortex-dominated flows because the spatial distribution of the vorticity-induced forces can be visualized, enabling detailed dissections of aerodynamic loads generated by individual vortical structures. For two-dimensional airfoils, \citet{menon2021significance} applied FMPM and showed that the strain-dominated region surrounding the rotation-dominated vortices has an important role to play in the generation of unsteady aerodynamic forces. For three-dimensional wings, this method has been implemented to study the contributions of spanwise and cross-span vortices to the lift generation of rectangular wings \citep{menon2022contribution}, the vorticity-induced force distributions on forward- and backward-swept wings at a fixed angle of attack \citep{zhang2022laminar}, and the aerodynamic forces on delta wings \citep{li2020vortex3d}. More recently, efforts have been made to apply FMPM to the analysis of experimental data, in particular, flow fields obtained using particle image velocimetry. \citet{zhu2023force} employed FMPM to analyze the vortex dynamics of a two-dimensional wing pitching sinusoidally in a quiescent flow. Several practical issues in applying FMPM to PIV data were discussed, including the effect of phase-averaging and potential error sources. 

In this study, we apply FMPM to three-dimensional flow field data measured using three-component PIV, and use the results to gain insight into the three-dimensional vortex dynamics and the corresponding unsteady forces acting on elastically mounted pitching swept wings. We extend the methodology developed in \citet{zhu2020nonlinear}, and employ a layered stereoscopic PIV technique and the FMPM to quantify the three-dimensional vortex dynamics. In the following sections, we first introduce the experimental setup and method of analysis (\S \ref{sec.setup}). The static force and moment coefficients of the wings are measured (\S \ref{sec.static}) before we characterize the amplitude response (\S \ref{sec.bifurcation}) and the frequency response (\S \ref{sec.frequency}) of the system. Next, we associate the onset of flow-induced oscillations with the static characteristics of the wing (\S \ref{sec.onset}) and use an energy approach to explain the nonlinear stability boundaries (\S \ref{sec.energy}). The unsteady force and moment measurements, together with the three-dimensional flow structures (\S \ref{sec.moment_PIV}) are then analyzed to explain the differences in power extraction for unswept and swept wings. Finally, we apply the Force and Moment Partitioning Method to quantitatively correlate the three-dimensional vortex dynamics with the resultant unsteady aerodynamic moment (\S \ref{sec.FMPM}). All the key findings are summarized in \S \ref{sec.conclusion}.

\vspace{1.5em}
\section{Methods}\label{sec.setup}

\begin{figure}
\centering
\includegraphics[width=0.9\textwidth]{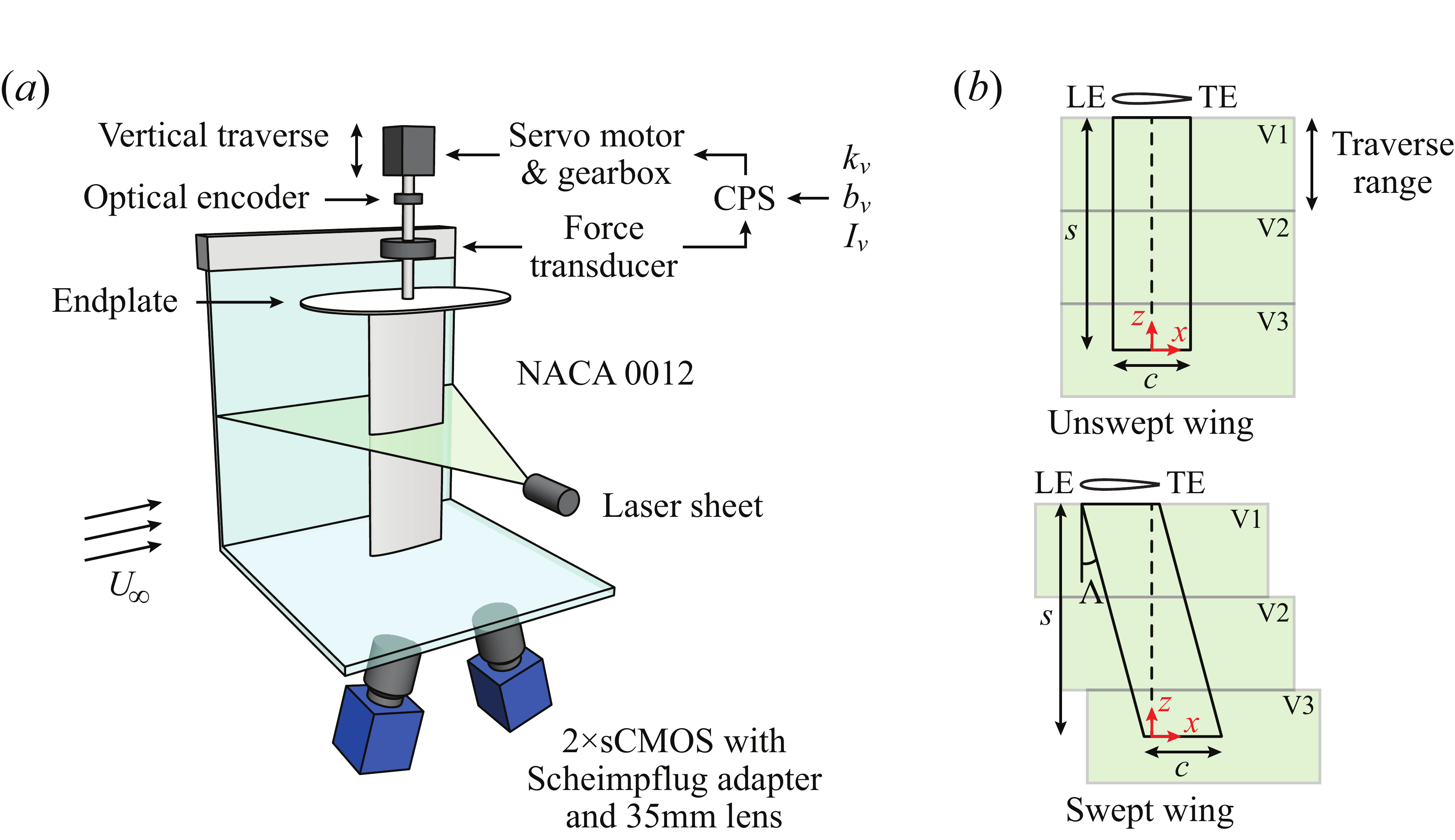}
\caption{(\emph{a}) A schematic of the experimental setup. (\emph{b}) Sketches of unswept and swept wings used in the experiments. The pivot axes are indicated by black dashed lines. The green panels represent volumes traversed by the laser sheet for three-dimensional phase-averaged stereoscopic PIV measurements.}
\label{fig.setup}
\end{figure}
\subsection{Cyber-physical system and wing geometry}{\label{sec.CPS}}

We perform all the experiments in the Brown University free-surface water tunnel, which has a test section of $W \times D \times L = 0.8~\mathrm{m} \times 0.6~\mathrm{m} \times 4.0~\mathrm{m}$. The turbulence intensity in the water tunnel is around 2\% at the velocity range tested in the present study. Free-stream turbulence plays a critical role in shaping small-amplitude laminar separation flutter (see \citet{yuan2015effect}). However, as we will show later, the flow-induced oscillations and the flow structures observed in the present study are of high amplitude and large size, and we do not expect the free-stream turbulence to play any significant role. Figure \ref{fig.setup}(\emph{a}) shows a schematic of the experimental setup. Unswept and swept NACA 0012 wings are mounted vertically in the tunnel, with an endplate on the top as a symmetry plane. The wing tip at the bottom does not have an endplate. The wings are connected to a six-axis force/moment transducer (ATI Delta IP65) via a wing shaft. The shaft further connects the transducer to an optical encoder (US Digital E3-2500) and a servo motor (Parker SM233AE) coupled with a gearbox (SureGear PGCN23-0525). 

We implement a cyber-physical system (CPS) to facilitate a wide structural parameter sweep (i.e. stiffness, $k$, damping, $b$, and inertia, $I$) while simulating real aeroelastic systems with high fidelity. Details of the CPS have been discussed in \citet{zhu2020nonlinear}, therefore, only a brief introduction will be given here. In the CPS, the force/moment transducer measures the fluid moment, $M$, and feeds the value to the computer via a data acquisition (DAQ) board (National Instruments PCIe-6353). This fluid moment is then added to the stiffness moment ($k\theta$) and the damping moment ($b\dot{\theta}$) obtained from the previous time step to get the total moment. Next, we divide this total moment by the desired inertia ($I$) to get the acceleration ($\ddot{\theta}$) at the present time step. This acceleration is then integrated once to get the velocity ($\dot{\theta}$) and twice to get the pitching angle ($\theta$). This pitching angle signal is output to the servo motor via the same DAQ board. The optical encoder, which is independent of the CPS, is used to measure and verify the pitching angle. At the next time step, the CPS recalculates the total moment based on the measured fluid moment and the desired stiffness and damping, and thereby continues the loop.

Our CPS control loop runs at a frequency of 4000 Hz, which is well beyond the highest Nyquist frequency of the aeroelastic system. Noise in the force/moment measurements can be a potential issue for the CPS. However, because we are using a position control loop, where the acceleration is integrated twice to get the desired position, our system is less susceptive to noise. Therefore, no filter is used within the CPS control loop. The position control loop also requires the pitching motor to follow the commanded position signal as closely as possible. This is achieved by carefully tuning the PID (Proportional–Integral–Derivative) parameters of the pitching motor. The CPS does not rely on any additional tunable parameters other than the virtual inertia, damping, and stiffness. We validate the system using `ring-down' experiments, as shown in the appendix of \citet{zhu2020nonlinear}. Moreover, as we will show later, the CPS results match remarkably well with prescribed experiments (\S\ref{sec.energy}), demonstrating the robustness of the system.

The unswept and swept wings used in the present study are sketched in figure \ref{fig.setup}(\emph{b}). All the wings have a span of $s=0.3$ m and a chord length of $c=0.1$ m, which results in a physical aspect ratio of $AR=3$. However, the effective aspect ratio is 6 due to the existence of the symmetry plane (i.e. the endplate). The minimum distance between the wing tip and the bottom of the water tunnel is around $1.5c$. The chord-based Reynolds number is defined as $Re \equiv \rho U_{\infty} c / \mu$, where $U_{\infty}$ is the free-stream velocity, $\rho$ and $\mu$ are water density and dynamic viscosity, respectively. We set the free-stream velocity to be $U_{\infty}=0.5$ $\mathrm{m~s^{-1}}$ for all the experiments (except for particle image velocimetry measurements, see \S \ref{sec.PIV_setup}), which results in a constant Reynolds number of $Re = 50~000$, matching the $Re$ used in \citet{zhu2020nonlinear} to facilitate direct comparisons. For both unswept and swept wings, the leading edge (LE) and the trailing edge (TE) are parallel. Their pivot axes, represented by vertical dashed lines in the figure, pass through the mid-chord point $x/c=0.5$ of the mid-span plane $z/s=0.5$. We choose the current location of the pitching axis because it splits the swept wings into two equal-area sections (fore and aft). Moving the pitching axis or making it parallel to the leading edge will presumably result in different system dynamics, which will be investigated in future studies.

The sweep angle, $\Lambda$, is defined as the angle between the leading edge and the vertical axis. Five wings with $\Lambda = 0^\circ$ (unswept wing), $10^\circ, 15^\circ, 20^\circ$ and $25^\circ$ (swept wings) are used in the experiments. Further expanding the range of wing sweep would presumably bring more interesting fluid-structure interaction behaviors. However, as we will show in the later sections, there is already a series of rich (nonlinear) flow physics associated with the current set of unswept and swept wings. Our selection of the sweep angle is also closely related to the location of the pitching axis. Currently, the pitching axis passes the mid-chord at the mid-span. For a $\Lambda=25^\circ$ wing, the trailing edge is already in front of the pitching axis at the wing root, and the leading edge is behind the pitching axis at the wing tip. Further increasing the sweep angle brings difficulties in physically pitching the wing for our existing setup.

\subsection{Multi-layer stereoscopic particle image velocimetry}{\label{sec.PIV_setup}}

We use multi-layer phase-averaged stereoscopic particle image velocimetry (SPIV) to measure the three-dimensional (3D) velocity field around the pitching wings. We lower the free-stream velocity to $U_{\infty}=0.3$ $\mathrm{m~s^{-1}}$ to enable higher temporal measurement resolution. The chord-based Reynolds number is consequently decreased to $Re = 30~000$. It has been shown by \citet[][see their appendix]{zhu2020nonlinear} that the variation of $Re$ in the range of 30~000 -- 60~000 does not affect the system dynamics, as long as the parameters of interest are properly non-dimensionalized. The water flow is seeded using neutrally buoyant 50 $\mu$m silver-coated hollow ceramic spheres (Potters Industries) and illuminated using a horizontal laser sheet, generated by a double-pulse Nd:YAG laser (532 nm, Quantel EverGreen) with a LaVision laser guiding arm and collimator. Two sCMOS cameras (LaVision, $2560 \times 2160$ pixels) with Scheimpflug adapters (LaVision) and 35mm lenses (Nikon) are used to capture image pairs of the flow field. These SPIV image pairs are fed into the LaVision DaVis software (v.10) for velocity vector calculation using multi-pass cross-correlations (two passes at $64 \times 64$ pixels, two passes at $32 \times 32$ pixels, both with 50\% overlap).

To measure the two-dimensional-three-component (2D3C) velocity field at different spanwise layers, we use a motorized vertical traverse system with a range of 120 mm to raise and lower the testing rig (i.e. all the components connected by the shaft) in the $z$-axis \citep{king2018experimental,zhong2021aspect}. Due to the limitation of the traversing range, three measurement volumes (figure \ref{fig.setup}\emph{b}, V1, V2 and V3) are needed to cover the entire wing span plus the wing tip region. For each measurement volume, the laser sheet is fixed at the top layer and the rig is traversed upward with a step size of 5 mm. Note that the entire wing stays submerged, even at the highest traversing position, and for all wing positions, free surface effects are not observed. The top two layers of V1 are discarded as the laser sheet is too close to the endplate, which causes reflections. The bottom layer of V1 and the top layer of V2 overlap with each other. The velocity fields of these two layers are averaged to smooth the interface between the two volumes. The interface between V2 and V3 is also smoothed in the same way. For each measurement layer, we phase-average 1250 instantaneously measured 2D3C velocity fields over 25 cycles (i.e. 50 measurements per cycle) to eliminate any instantaneous variations of the flow field while maintaining the key coherent features across different layers. Finally, 71 layers of 2D3C velocity fields are stacked together to form a large volume of phase-averaged 3D3C velocity field ($\sim3c \times 3c \times 3.5c$). The velocity fields of three wing models ($\Lambda=0^\circ$, $10^\circ$ and $20^\circ$) are measured. For the two swept wings ($\Lambda=10^\circ$ and $20^\circ$), the laser volumes are offset horizontally to compensate for the sweep angle (see the bottom subfigure of figure \ref{fig.setup}\emph{b}).


\subsection{Governing equations and non-dimensional parameters}{\label{sec.equations}}

The one-degree-of-freedom aeroelastic system considered in the present study has a governing equation
\begin{equation}
    I \ddot{\theta} + b \dot{\theta} + k \theta = M,
\label{eqn.govern}
\end{equation}
where $\theta$, $\dot{\theta}$, and $\ddot{\theta}$ are the angular position, velocity and acceleration, respectively. $I=I_p+I_v$ is the effective inertia, where $I_p$ is the physical inertia of the wing and $I_v$ is the virtual inertia that we prescribe with the CPS. Because the friction is negligible in our system, the effective structural damping, $b$, equals the virtual damping $b_v$ in the CPS. $k$ is the effective torsional stiffness and it equals the virtual stiffness $k_v$. Equation \ref{eqn.govern} resembles a forced torsional spring-mass-damper system, where the fluid moment, $M$, acts as a nonlinear forcing term. Following \citet{onoue2015large} and \citet{zhu2020nonlinear}, we normalize the effective inertia, damping, stiffness and the fluid moment using the fluid inertia force to get the non-dimensional governing equation of the system:
\begin{equation}
    I^* \ddot{\theta}^* + b^* \dot{\theta}^* + k^* \theta^* = C_M,
\label{eqn.normlized_govern}
\end{equation}
where
\begin{equation}\label{eqn.parameters}
\begin{gathered}
    \theta^* = \theta,~\dot{\theta}^*=\frac{\dot{\theta}c}{U_{\infty}},~\ddot{\theta}^* = \frac{\ddot{\theta}c^2}{U_{\infty}^2},\\
    I^* = \frac{I}{0.5\rho c^4 s},~b^* = \frac{b}{0.5\rho U_{\infty} c^3 s},~k^* = \frac{k}{0.5\rho U_{\infty}^2 c^2 s},~C_M = \frac{M}{0.5\rho U_{\infty}^2 c^2 s}.    
\end{gathered}
\end{equation}
We should note that the inverse of the non-dimensional stiffness is equivalent to the Cauchy number, $Ca=1/k^*$, and the non-dimensional inertia, $I^*$, is analogous to the mass ratio between the wing and the surrounding fluid. We define the non-dimensional velocity as $U^*=U_{\infty}/(2 \pi f_p c)$, where $f_p$ is the \emph{measured} pitching frequency. In addition to the aerodynamic moment, we also measure the aerodynamic forces that are normal and tangential to the wing chord, $F_N$ and $F_T$, respectively. The resultant lift and drag forces are
\begin{equation}\label{eqn.lift_drag}
\begin{gathered}
    L = F_N \cos{\theta} - F_T \sin{\theta}, \\
    D = F_N \sin{\theta} + F_T \cos{\theta}.    
\end{gathered}
\end{equation}
We further normalize the normal force, tangential force, lift and drag to get the corresponding force coefficients
\begin{equation}
    C_N = \frac{F_N}{0.5\rho U_{\infty}^2 c s}, ~C_T = \frac{F_T}{0.5\rho U_{\infty}^2 c s},~C_L = \frac{L}{0.5\rho U_{\infty}^2 c s}, ~C_D = \frac{D}{0.5\rho U_{\infty}^2 c s}.
\end{equation}

\subsection{Force and Moment Partitioning Method}{\label{sec.FMPM_intro}}

To apply FMPM to three-dimensional PIV data, we first construct an influence potential that satisfies Laplace's equation and two different Neumann boundary conditions on the airfoil and the outer boundary
\begin{equation}
    \nabla^2 \phi = 0,~\text{and}~\frac{\partial \phi}{\partial \boldsymbol{\mathrm{n}}} =
    \begin{cases}
      [(\boldsymbol{x}-\boldsymbol{x_p})\times\boldsymbol{\mathrm{n}}]\cdot\boldsymbol{\mathrm{e_z}} & \text{on airfoil} \\
      0 & \text{on outer boundary}
    \end{cases},
\label{eqn.3d_phi}
\end{equation}
where $\boldsymbol{\mathrm{n}}$ is the unit vector normal to the boundary, $\boldsymbol{x}-\boldsymbol{x_p}$ is the location vector pointing from the pitching axis $\boldsymbol{x_p}$ towards location $\boldsymbol{x}$ on the airfoil surface, and $\boldsymbol{\mathrm{e_z}}$ is the spanwise unit vector \citep{menon2021quantitative}. This influence potential quantifies the spatial influence of any vorticity on the resultant force/moment. It is only a function of the airfoil geometry and the pitching axis, and does not depend on the kinematics of the wing. Note that this influence potential should not be confused with the velocity potential from the potential flow theory. The boundary conditions of equation \ref{eqn.3d_phi} are specified for solving the influence field of the spanwise moment, and they will be different for solving the lift and drag influence fields. From the three-dimensional velocity data, we can calculate the $Q$ field \citep{hunt1988eddies,jeong1995identification}
\begin{equation}
Q=\frac{1}{2}(\Vert\boldsymbol{\Omega}\Vert^2-\Vert\boldsymbol{\mathrm{S}}\Vert^2), 
\label{eq:Q-def}
\end{equation}
where $Q$ is the second invariant of the velocity gradient tensor, $\boldsymbol{\Omega}$ is the vorticity tensor and $\boldsymbol{\mathrm{S}}$ is the strain-rate tensor. The vorticity-induced moment can be evaluated by 
\begin{equation}
    M_v = -2\rho \int_V Q \phi~\mathrm{d}V,
\label{eqn.vortex_moment}
\end{equation}
where $\int_V$ represents the volume integral within the measurement volume. The spatial distribution of the vorticity-induced moment near the pitching wing can thus be represented by the moment density, $-2Q\phi$ (i.e. the moment distribution field). In the present study, we focus on the vorticity-induced force (moment) as it has the most important contribution to the overall unsteady aerodynamic load in vortex-dominated flows. Other forces including the added-mass force, the force due to viscous diffusion, the forces associated with irrotational effects and outer domain effects are not considered although they can be estimated using FMPM as well \citep{menon2021quantitative}. The contributions from these other forces, along with experimental errors, might result in a mismatch in the magnitude of the FMPM-estimated force and force transducer measurements, as shown by \citet{zhu2023force}, and the exact source of this mismatch is under investigation.

\section{Results and discussion}
\subsection{Static characteristics of unswept and swept wings}{\label{sec.static}}

\begin{figure}
\centering
\includegraphics[width=1\textwidth]{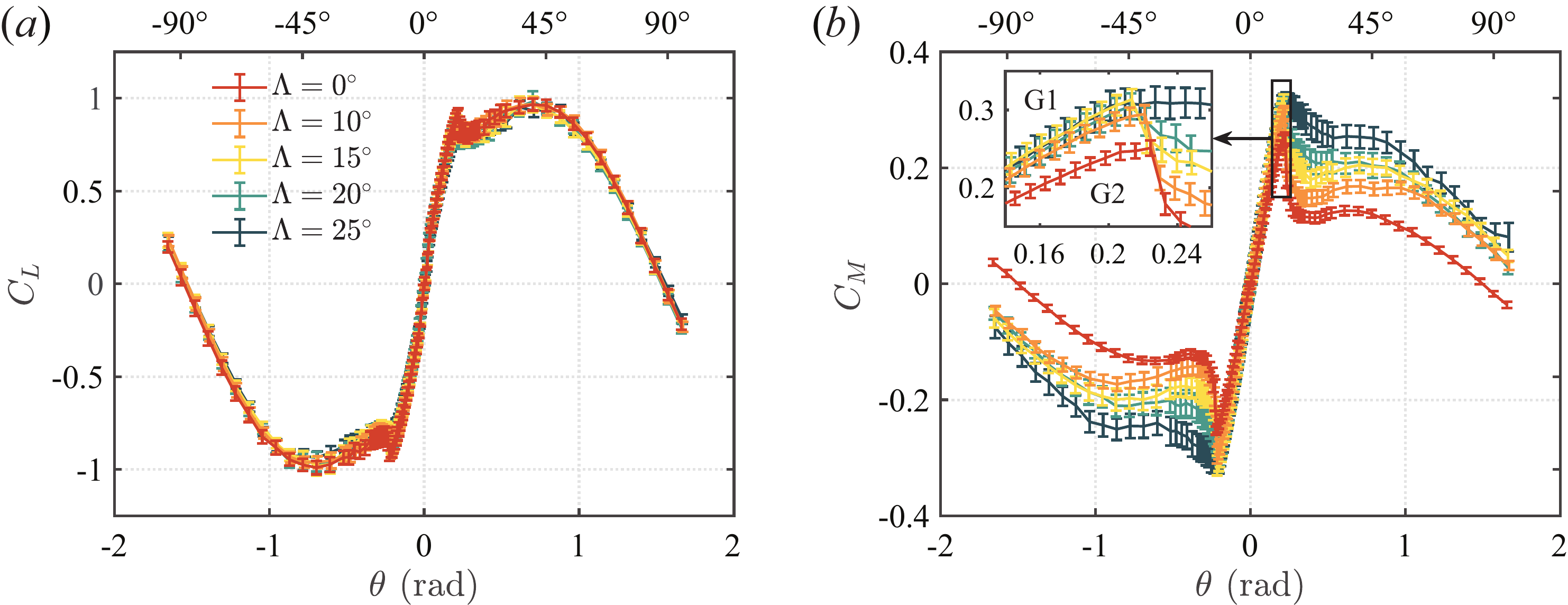}
\caption{(\emph{a}) Static lift coefficient and (\emph{b}) moment coefficient of unswept and swept wings. Error bars denote standard deviations of the measurement over 20 seconds.}
\label{fig.static}
\end{figure}

The static lift and moment coefficient, $C_L$ and $C_M$, are measured for the unswept ($\Lambda = 0^\circ$) and swept wings ($\Lambda = 10^\circ$ -- $25^\circ$) at $Re = 50~000$ and the results are shown in figure \ref{fig.static}. In figure \ref{fig.static}(\emph{a}), we see that the static lift coefficient, $C_L(\theta)$, has the same behavior for all sweep angles, despite some minor variations for angles of attack higher than the static stall angle $\theta_s = 12^\circ$ (0.21 rad). The collapse of $C_L(\theta)$ across different swept wings agrees with the classic `independence principle' \citep{jones1947effects} (i.e. $C_L \sim \cos^2\Lambda$) at relatively small sweep angles. Figure \ref{fig.static}(\emph{b}) shows that, for any fixed angle of attack, the static moment coefficient, $C_M$, increases with the sweep angle, $\Lambda$. This trend is most prominent when the angle of attack exceeds the static stall angle. The inset shows a zoom-in view of the static $C_M$ for $\theta = 0.14$ -- 0.26. It is seen that the $C_M$ curves cluster into two groups, with the unswept wing ($\Lambda = 0^\circ$) being in Group 2 (G2) and all the other swept wings ($\Lambda = 10^\circ$ -- $25^\circ$) being in Group 1 (G1). As we will show later, this grouping behavior is closely related to the onset of flow-induced oscillations (\S \ref{sec.bifurcation} \& \S \ref{sec.onset}) and it is important for understanding the system stability. No hysteresis is observed for both static $C_L$ and $C_M$, presumably due to free-stream turbulence in the water tunnel.

\subsection{Subcritical bifurcations to flow-induced oscillations}{\label{sec.bifurcation}}

We conduct bifurcation tests to study the stability boundaries of the elastically mounted pitching wings. \citet{zhu2020nonlinear} have shown that for unswept wings, the onset of limit-cycle oscillations (LCOs) is independent of the wing inertia and the bifurcation type (i.e. subcritical or supercritical). It has also been shown that the extinction of LCOs for subcritical bifurcations at different wing inertias occurs at a fixed value of the non-dimensional velocity $U^*$. For these reasons, we choose to focus on one high-inertia case ($I^*=10.6$) in the present study. In the experiments, the free-stream velocity is maintained at $U_\infty = 0.5$ $\mathrm{m~s^{-1}}$. We fix the structural damping of the system at a small value, $b^*=0.13$, keep the initial angle of attack at zero, and use the Cauchy number, $Ca$, as the control parameter. To test for the onset of LCOs, we begin the test with a high-stiffness virtual spring (i.e. low $Ca$) and incrementally increase $Ca$ by decreasing the torsional stiffness, $k^*$. We then reverse the operation to test for the extinction of LCOs and to check for any hysteresis. The amplitude response of the system, $A$, is measured as the peak absolute pitching angle (averaged over many pitching cycles). By this definition, $A$ is half of the peak-to-peak amplitude. The divergence angle, $\overline{A}$, is defined as the mean absolute pitching angle. Although all the divergence angles are shown to be positive, the wing can diverge to both positive and negative angles in experiments.

\begin{figure}
\centering
\includegraphics[width=.6\textwidth]{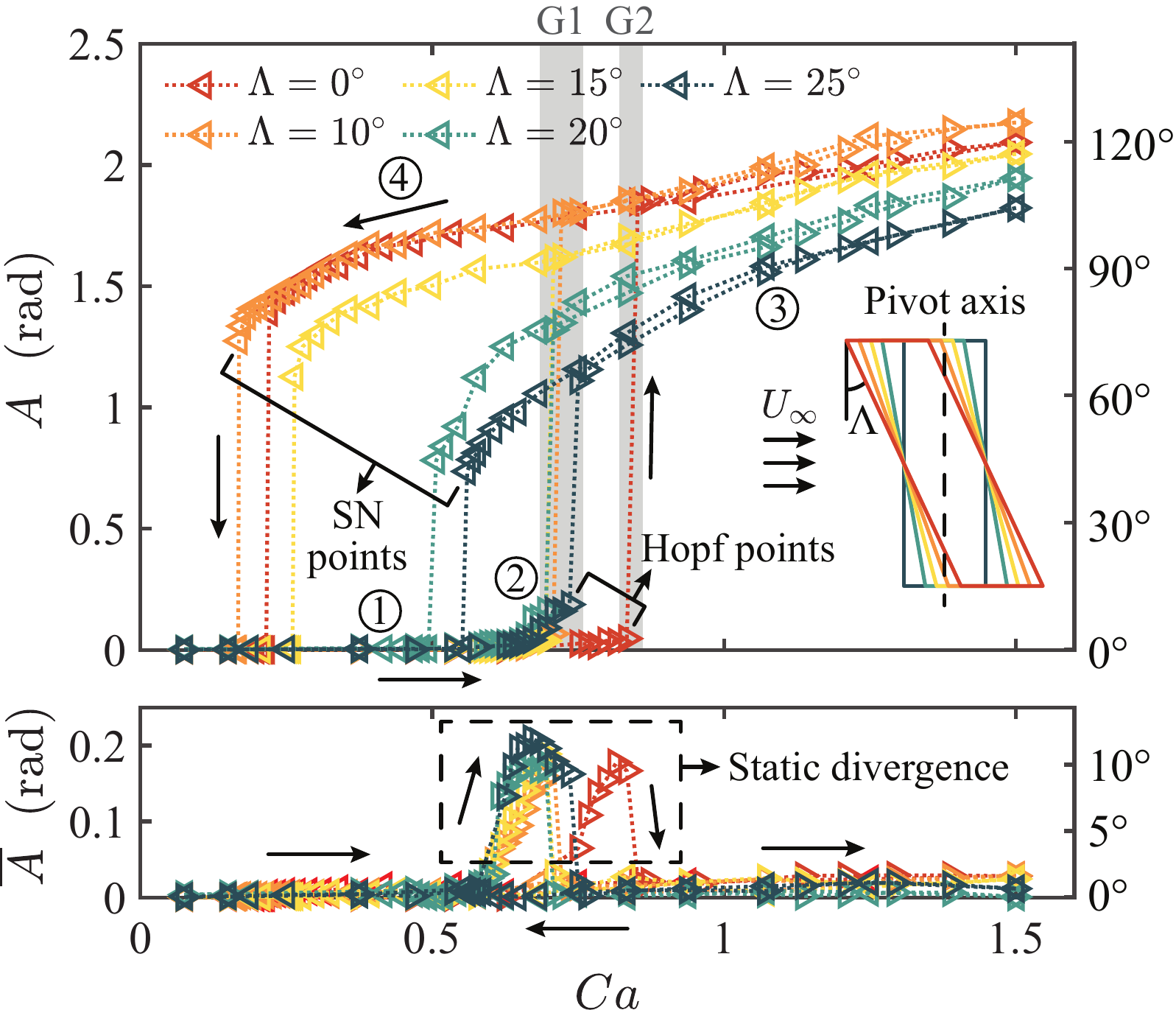}
\caption{Amplitude response and static divergence for unswept and swept wings. $\triangleright$: increasing $Ca$, $\triangleleft$: decreasing $Ca$. The inset illustrates the wing geometry and the pivot axis. The colors of the wings correspond to the colors of the amplitude and divergence curves in the figure.}
\label{fig.bifurcation}
\end{figure}

Figure \ref{fig.bifurcation} shows the pitching amplitude response and the static divergence angle for swept wings with $\Lambda=10^\circ$ to $25^\circ$. Data for the unswept wing ($\Lambda=0^\circ$) are also replotted from \citet{zhu2020nonlinear} for comparison. It can be seen that the system first remains stable without any noticeable oscillations or divergence (regime \circled{1} in the figure) when $Ca$ is small. In this regime, the high stiffness of the system is able to pull the system back to a stable fixed point despite any small perturbations. As we further increase $Ca$, the system diverges to a small static angle, where the fluid moment is balanced by the virtual spring. This transition is presumably triggered by free-stream turbulence, and both positive and negative directions are possible. Due to the existence of random flow disturbances and the decreasing spring stiffness, some small-amplitude oscillations around the static divergence angle start to emerge (regime \circled{2}). As $Ca$ is further increased above a critical value (i.e. the Hopf point), the amplitude response of the system abruptly jumps into large-amplitude self-sustained LCOs and the static divergence angle drops back to zero, indicating that the oscillations are symmetric about the zero angle of attack. The large-amplitude LCOs are observed to be near-sinusoidal and have a dominant characteristic frequency. After the bifurcation, the amplitude response of the system continues to increase with $Ca$ (regime \circled{3}). We then decrease $Ca$ and find that the large-amplitude LCOs persist even when $Ca$ is decreased below the Hopf point (regime \circled{4}). Finally, the system drops back to the stable fixed point regime via a saddle-node (SN) point. A hysteretic bistable region is thus created in between the Hopf point and the saddle-node point -- a hallmark of a subcritical Hopf bifurcation. In the bistable region, the system features two stable solutions -- a stable fixed point (regime \circled{1}) and a stable LCO (regime \circled{4}) -- as well as an unstable LCO solution, which is not observable in experiments \citep{strogatz1994nonlinear}.

We observe that the Hopf points of unswept and swept wings can be roughly divided into two groups (figure \ref{fig.bifurcation}, G1 \& G2), with the unswept wing ($\Lambda=0^\circ$) being in G2 and all the other swept wings ($\Lambda=10^\circ$ -- $25^\circ$) being in G1, which agrees with the trend observed in figure \ref{fig.static}(\emph{b}) for the static moment coefficient. This connection will be discussed further in \S \ref{sec.onset}. It is also seen that as the sweep angle increases, the LCO amplitude at the saddle-node point decreases monotonically. However, the $Ca$ at which the saddle-node point occurs first extends towards a lower value ($\Lambda=0^\circ \rightarrow 10^\circ$) but then moves back towards a higher $Ca$ ($\Lambda=10^\circ \rightarrow 25^\circ$). This indicates that increasing the sweep angle first destabilizes the system from $\Lambda=0^\circ$ to $10^\circ$ and then re-stabilizes it from $\Lambda=10^\circ$ to $25^\circ$. This non-monotonic behavior of the saddle-node point will be revisited from a perspective of energy in \S \ref{sec.energy}. The pitching amplitude response, $A$, follows a similar non-monotonic trend. Between $\Lambda=0^\circ$ and $10^\circ$, $A$ is slightly higher at higher $Ca$ values for the $\Lambda=10^\circ$ wing, whereas between $\Lambda=10^\circ$ and $25^\circ$, $A$ decreases monotonically, indicating that a higher sweep angle is not able to sustain LCOs at higher amplitudes. The non-monotonic behaviors of the saddle-node point and the LCO amplitude both suggest that there exists an optimal sweep angle, $\Lambda=10^\circ$, which promotes flow-induced oscillations of pitching swept wings.

\subsection{Frequency response of the system}{\label{sec.frequency}}

\begin{figure}
\centering
\includegraphics[width=1\textwidth]{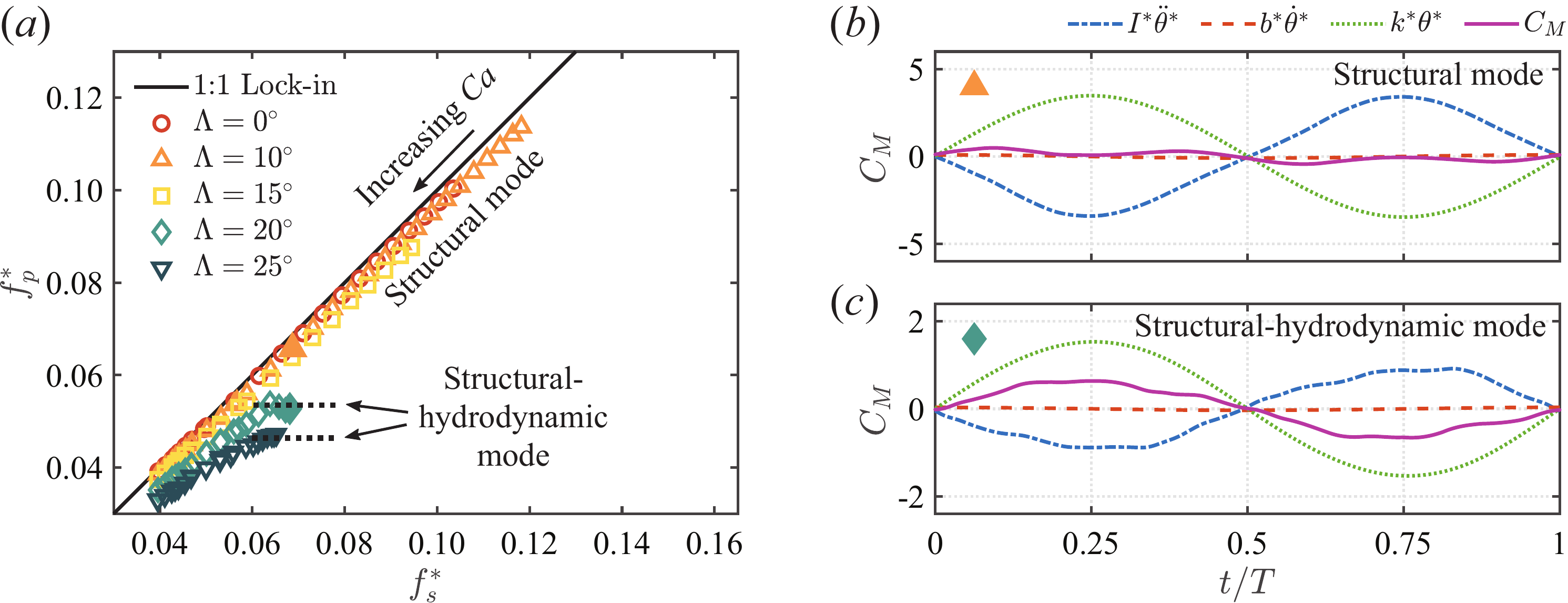}
\caption{(\emph{a}) Frequency response of unswept and swept wings. (\emph{b}, \emph{c}) Force decomposition of the structural mode and the structural-hydrodynamic mode. (\emph{b}) and (\emph{c}) correspond to the filled orange triangle and the filled green diamond shown in (\emph{a}), respectively. Note that $t/T=0$ corresponds to $\theta = 0$.}
\label{fig.frequency}
\end{figure}

The characteristic frequencies of the flow-induced LCOs observed in figure \ref{fig.bifurcation} provide us with more information about the driving mechanism of the oscillations. Figure \ref{fig.frequency}(\emph{a}) shows the measured frequency response, $f_p^*$, as a function of the calculated natural (structural) frequency, $f_s^*$, and sweep angle. In the figure, $f_p^* = f_p c /U_{\infty}$ and $f_s^* = f_s c /U_{\infty}$, where $f_p$ is the measured pitching frequency and
\begin{equation}
    f_s = \frac{1}{2\pi} \sqrt{\frac{k}{I}-(\frac{b}{2I})^2}
\label{eqn.frequency}
\end{equation}
is the structural frequency of the system \citep{rao1995mechanical}. We observe that for all the wings tested in the experiments and over most of the regimes tested, the measured pitching frequency, $f_p^*$, locks onto the calculated structural frequency, $f_s^*$, indicating that the oscillations are dominated by the balance between the structural stiffness and inertia. These oscillations, therefore, correspond to the \emph{structural} mode reported by \citet{zhu2020nonlinear}, and feature characteristics of high-inertial aeroelastic instabilities. We can decompose the moments experienced by the wing into the inertial moment, $I^* \ddot{\theta}^*$, the structural damping moment, $b^* \dot{\theta}^*$, the stiffness moment, $k^* \theta^*$, and the fluid moment, $C_M$. As an example, for the $\Lambda=10^\circ$ wing pitching at $f^*_s=0.069$ (i.e. the filled orange triangle in figure \ref{fig.frequency}\emph{a}), these moments are plotted in figure \ref{fig.frequency}(\emph{b}). We see that for the structural mode, the stiffness moment is mainly balanced by the inertial moment, while the structural damping moment and the fluid moment remain relatively small. 

In addition to the structural mode, \citet{zhu2020nonlinear} also observed a hydrodynamic mode, which corresponds to a low-inertia wing. In the hydrodynamic mode, the oscillations are dominated by the fluid forcing, so that the measured pitching frequency, $f^*_p$, stays relatively constant for a varying $Ca$. In figure \ref{fig.frequency}(\emph{a}), we see that for the $\Lambda=20^\circ$ and $25^\circ$ wings, $f^*_p$ flattens near the saddle-node boundary. This flattening trend shows an emerging fluid-dominated time scale, resembling a hydrodynamic mode despite the high wing inertia. Taking $\Lambda=20^\circ$, $f^*_s=0.068$ (i.e. the filled green diamond in figure \ref{fig.frequency}\emph{a}) as an example, we can examine the different contributions to the pitching moments in figure \ref{fig.frequency}(\emph{c}). It is observed that in this oscillation mode, the stiffness moment balances both the inertial moment and the fluid moment. This is different from both the structural mode and the hydrodynamic mode, and for this reason, we define this hybrid oscillation mode as the \emph{structural-hydrodynamic} mode. 

There are currently no quantitative descriptions of the structural-hydrodynamic mode. However, it can be qualitatively identified as when the pitching frequency of a (1:1 lock-in) structural mode flattens as the natural (structural) frequency increases. Based on the observations in the present study, we believe this mode is not a fixed fraction of the structural frequency. Instead, the frequency response shows a mostly flat trend (figure \ref{fig.frequency}\emph{a}, green and dark green curves) at high $f_s^*$, indicating an increasingly dominating fluid forcing frequency. For a structural mode, the oscillation frequency locks onto the natural frequency due to the high inertial moment. However, as the sweep angle increases, the fluid moment also increases (see also figure \ref{fig.PIV_traj}\emph{a}). The structural-hydrodynamic mode emerges as the fluid forcing term starts to dominate in the nonlinear oscillator.

For a fixed structural frequency, $f_s^*$, as the sweep angle increases, the measured pitching frequency, $f_p^*$, deviates from the 1:1 lock-in curve and moves to lower frequencies. This deviation suggests a growing added-mass effect, as the pitching frequency $f_p\sim\sqrt{1/(I+I_{add})}$. Because the structural inertia $I$ is prescribed, a decreasing $f_p$ suggests an increasing added-mass inertia, $I_{add}$. This is expected because of the way we pitch the wings in the experiments (see the inset of figure \ref{fig.bifurcation}). As $\Lambda$ increases, the accelerated fluid near the wing root and the wing tip produces more moments due to the increase of the moment arm, which amplifies the added-mass effect. The peak added-mass moment is estimated to be around 2\%, 3\%, and 5\% of the peak total moment for the $\Lambda=0^\circ$, $10^\circ$, and $20^\circ$ wings, respectively. Because this effect is small compared to the structural and vortex-induced forces, we will not quantify this added-mass effect further in the present study but will leave it for future work.

\subsection{Onset of flow-induced oscillations}{\label{sec.onset}}

\begin{figure}
\centering
\includegraphics[width=1\textwidth]{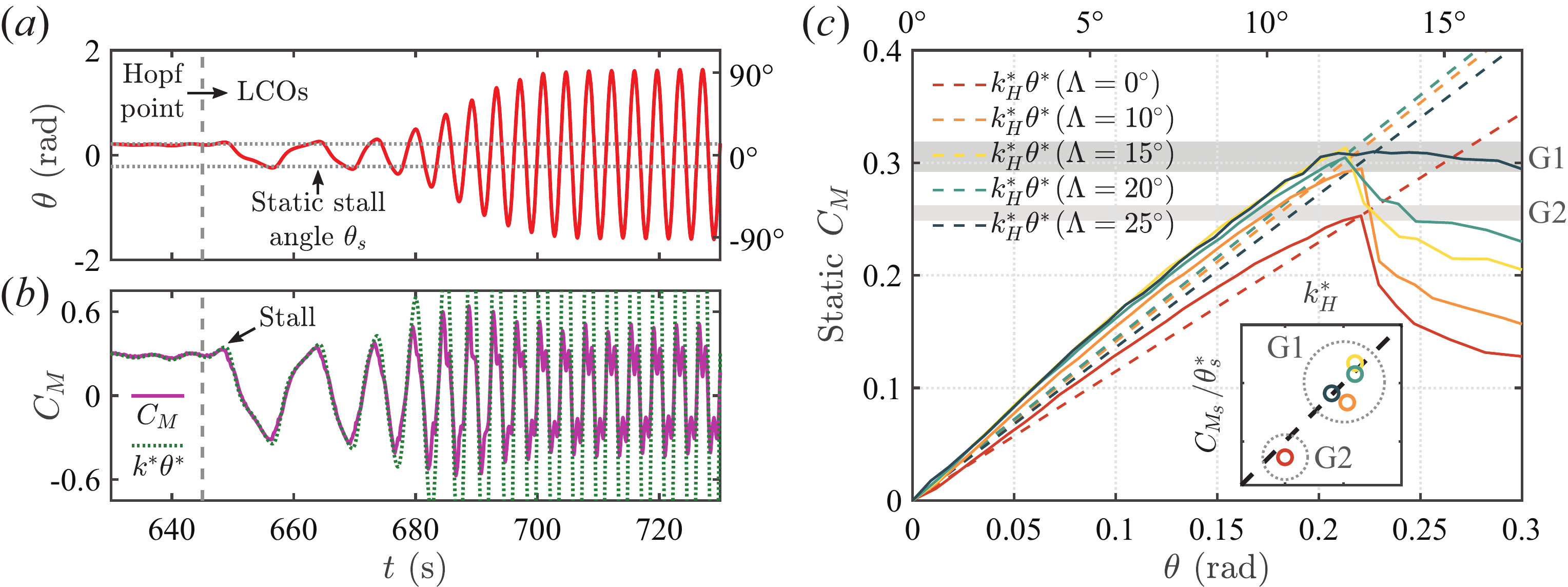}
\caption{Temporal evolution of (\emph{a}) the pitching angle $\theta$, (\emph{b}) the fluid moment $C_M$, and the stiffness moment $k^*\theta^*$ near the Hopf point for the $\Lambda=15^\circ$ swept wing. The vertical gray dashed line indicates the time instant ($t=645$ s) at which $Ca$ is increased above the Hopf point. (\emph{c}) Static moment coefficients of unswept and swept wings. Inset: The predicted Hopf point based on the static stall angle and the corresponding moment, $C_{M_s}/\theta_s^*$, versus the measured Hopf point, $k_H^*$. The black dashed line shows a 1:1 scaling.}
\label{fig.onset}
\end{figure}

In figure \ref{fig.bifurcation}, we have observed that the Hopf point of unswept and swept wings can be roughly divided into two groups (figure \ref{fig.bifurcation}, G1 \& G2). In this section, we explain this phenomenon. Figure \ref{fig.onset}(\emph{a}) and (\emph{b}) shows the temporal evolution of the pitching angle, $\theta(t)$, the fluid moment, $C_M(t)$, and the stiffness moment, $k^*\theta^*(t)$, for the $\Lambda=15^\circ$ swept wing as the Cauchy number is increased past the Hopf point. We see that the wing undergoes small amplitude oscillations around the divergence angle just prior to the Hopf point ($t<645$ s). The divergence angle is lower than the static stall angle, $\theta_s$, and so we know that the flow stays mostly attached, and the fluid moment, $C_M$, is balanced by the stiffness moment, $k^*\theta^*$ (figure \ref{fig.onset}\emph{b}). When the Cauchy number, $Ca = 1/k^*$, is increased above the Hopf point (figure \ref{fig.onset}\emph{a}, $t>645$ s), $k^* \theta^*$ is no longer able to hold the pitching angle below $\theta_s$. Once the pitching angle exceeds $\theta_s$, stall occurs and the wing experiences a sudden drop in $C_M$. The stiffness moment, $k^*\theta^*$, loses its counterpart and starts to accelerate the wing to pitch towards the opposite direction. This acceleration introduces unsteadiness to the system and the small-amplitude oscillations gradually transition to large-amplitude LCOs over the course of several cycles, until the inertial moment kicks in to balance $k^*\theta^*$ (see also figure \ref{fig.frequency}\emph{b}). This transition process confirms the fact that the onset of large-amplitude LCOs depends largely on the \emph{static} characteristics of the wing -- the LCOs are triggered when the static stall angle is exceeded.

The triggering of flow-induced LCOs starts from $\theta$ exceeding the static stall angle after $k^*$ is decreased below the Hopf point, causing $C_M$ to drop below $k^*\theta^*$. At this value of $k^*$, the slope of the static stall point should be equal to the stiffness at the Hopf point, $k^*_H$ (i.e. $C_{M_s} = k^*_H \theta^*$, where $C_{M_s}$ is the static stall moment). This argument is verified by figure \ref{fig.onset}(\emph{c}), in which we replot the static moment coefficients of unswept and swept wings from figure \ref{fig.static}(\emph{b}) (error bars omitted for clarity), together with the corresponding $k^*_H \theta^*$. We see that the $k^*_H \theta^*$ lines all roughly pass through the static stall points ($\theta_s^*$, $C_{M_s}$) of the corresponding $\Lambda$. Note that $k^*_H \theta^*$ of $\Lambda=15^\circ$ and $20^\circ$ overlap with each other. Similar to the trend observed for the Hopf point in figure \ref{fig.bifurcation}, the static stall moment $C_{M_s}$ can also be divided into two groups, with the unswept wing ($\Lambda=0^\circ$) being in G2 and all the other wings ($\Lambda=10^\circ$ -- $25^\circ$) being in G1 (see also figure \ref{fig.static}\emph{b}). The inset compares the predicted Hopf point, $C_{M_s}/\theta_s^*$, with the measured Hopf point, $k_H^*$, and we see that data closely follow a 1:1 relationship. This reinforces the argument that the onset of flow-induced LCOs is shaped by the static characteristics of the wing, and that this explanation applies to both unswept and swept wings.

It is worth noting that \citet{negi2021onset} performed global linear stability analysis on an aeroelastic wing and showed that the aeroelastic instability is triggered by a zero-frequency linear divergence mode. This agrees in part with our experimental observation that the flow-induced oscillations emerge from the static divergence state. However, as we have discussed in this section, the onset of large-amplitude aeroelastic oscillations in our system occurs when the divergence angle exceeds the static stall angle, whereas no stall is involved in the study of \citet{negi2021onset}. In fact, \citet{negi2021onset} focused on laminar separation flutter, where the pitching amplitude is small ($A\sim 6^\circ$). In contrast, we focus on large-amplitude ($45^\circ<A<120^\circ$) flow-induced oscillations.

\subsection{Power coefficient map and system stability}{\label{sec.energy}}

\begin{figure}
\centering
\includegraphics[width=0.9\textwidth]{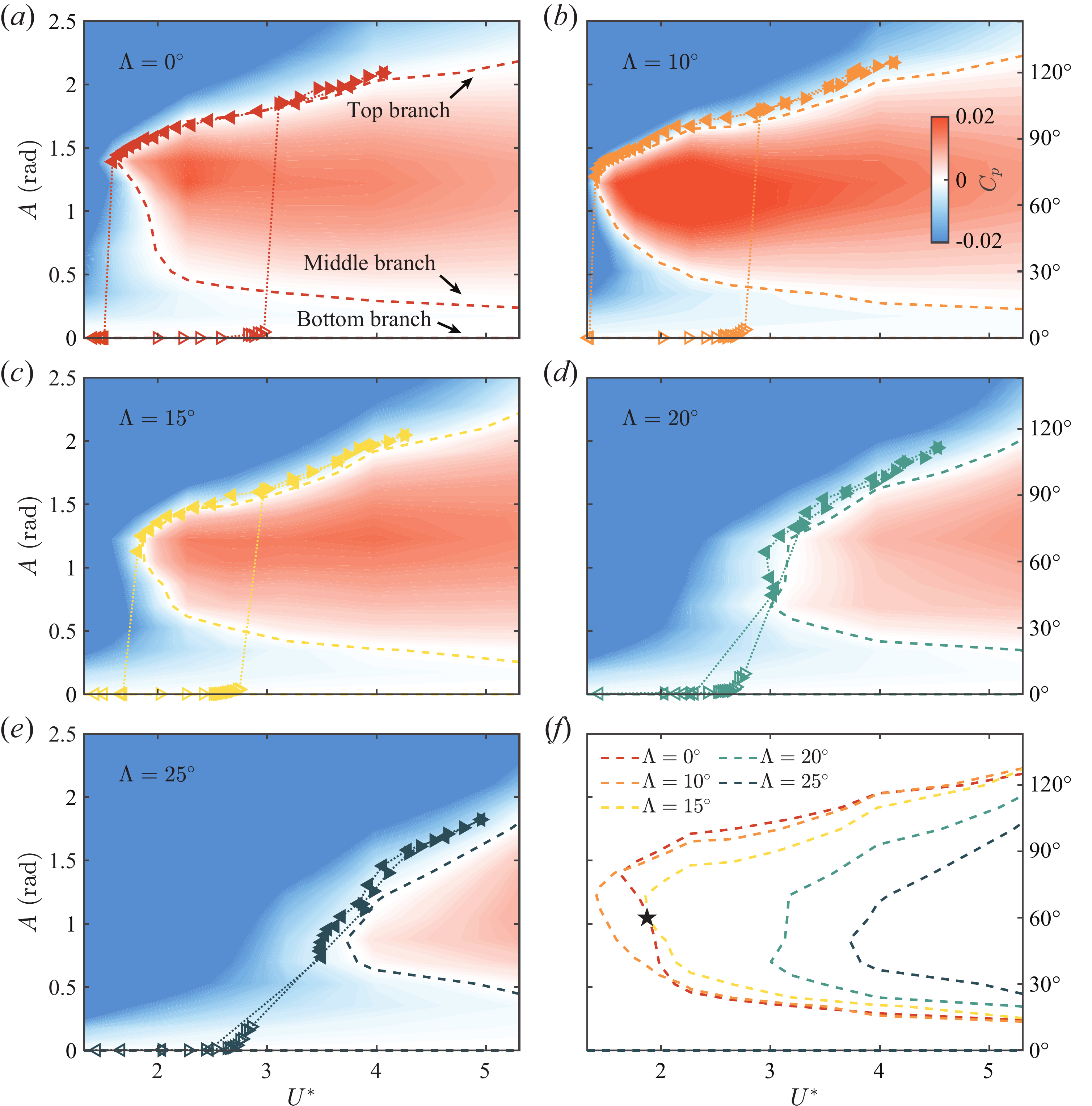}
\caption{(\emph{a}-\emph{e}) Power coefficient maps of prescribed sinusoidal oscillations overlaid by the bifurcation diagrams of elastically mounted unswept and swept wings. $\triangleright$: increasing $Ca$, $\triangleleft$: decreasing $Ca$. (\emph{f}) Neutral power transfer curves for unswept and swept wings. The black star represents the case $U^*=1.87$ ($f_p^*=0.085$), $A=1.05$ ($60^\circ$), where stereo PIV measurements are taken.}
\label{fig.energy}
\end{figure}

In this section, we analyze the stability of elastically mounted unswept and swept wings from the perspective of energy transfer. \citet{menon2019flow} and \citet{zhu2020nonlinear} have shown numerically and experimentally that the flow-induced oscillations of elastically mounted wings can only sustain when the net energy transfer between the ambient fluid and the elastic mount equals zero. To map out this energy transfer for a large range of pitching frequencies and amplitudes, we \emph{prescribe} the pitching motion of the wing using a sinusoidal profile
\begin{equation}
    \theta = A \sin (2\pi f_p t),
    \label{eqn.sine_motion}
\end{equation}
where $0 \leq A \leq 2.5$ rad and $0.15~\mathrm{Hz} \leq f_p \leq 0.6~\mathrm{Hz}$. The fluid moment $C_M$ measured with these prescribed sinusoidal motions can be directly correlated to those measured in the passive flow-induced oscillations because the flow-induced oscillations are near-sinusoidal (see \S \ref{sec.bifurcation}, and figure \ref{fig.onset}\emph{a}, $t>700$ s). By integrating the governing equation of the passive system \ref{eqn.normlized_govern} over $n=20$ cycles and taking the cycle average \citep{zhu2020nonlinear}, we can get the power coefficient of the system
\begin{equation}\label{eqn.power_balance}
    C_p = \frac{f_p^*}{n} \int_{t_0}^{t_0+nT} (C_M \dot{\theta}^* - b^* \dot{\theta}^{*2})~d t^*,
\end{equation}
where $t_0$ is the starting time, $T$ is the pitching period and $t^*=tU_{\infty}/c$ is the non-dimensional time. In this equation, the $C_M \dot{\theta}^*$ term represents the power injected into the system from the free-stream flow, whereas the $b^* \dot{\theta}^{*2}$ term represents the power dissipated by the structural damping of the elastic mount. The power coefficient maps of unswept and swept wings are shown in figure \ref{fig.energy}(\emph{a}-\emph{e}). In these maps, orange regions correspond to $C_p>0$, where the power injected by the ambient flow is higher than that dissipated by the structural damping. On the contrary, $C_p<0$ in the blue regions. The colored dashed lines indicate the $C_p=0$ contours, where the power injection balances the power dissipation, and the system is in equilibrium. The $C_p=0$ equilibrium boundary can be divided into three branches. \citet{zhu2020nonlinear} have shown that for unswept wings, the top branch corresponds to a stable LCO solution for the structural oscillation mode, the middle branch represents an unstable LCO solution for the structural mode, but a stable LCO solution for the hydrodynamic mode, and the bottom branch is a fixed point solution.

To correlate the power coefficient maps of prescribed oscillations with the stability boundaries of flow-induced oscillations, we overlay the bifurcation diagrams of the passive system from figure \ref{fig.bifurcation} onto figure \ref{fig.energy}(\emph{a}-\emph{e}). The measured pitching frequencies, $f_p$, are used to calculate the non-dimensional velocity, $U^*$, for large-amplitude LCOs (filled triangles). Because it is difficult to measure frequencies of fixed points and small-amplitude oscillations, we use the calculated structural frequency, $f_s$, to evaluate $U^*$ for non-LCO data points (hollow triangles). Figure \ref{fig.energy}(\emph{a}-\emph{e}) show that for all the wings tested, the flow-induced large-amplitude LCOs match well with the top branch of the $C_p=0$ curve, indicating the broad applicability of the energy approach for both unswept and swept wings, and confirming that this instability is a structural mode, as seen in the frequency response (figure \ref{fig.frequency}\emph{a}). This correspondence was also observed by \citet{menon2019flow} and \citet{zhu2020nonlinear} and is expected for instabilities that are well-described by sinusoidal motions \citep{morse2009prediction}. The small discrepancies for large sweep angles can be attributed to the low $C_p$ gradient near $C_p=0$. The junction between the top and the middle $C_p=0$ branches, which corresponds to the saddle-node point, stays relatively sharp for $\Lambda = 0^\circ$ -- $15^\circ$ and becomes smoother for $\Lambda = 20^\circ$ -- $25^\circ$. These smooth turnings result in a smooth transition from the structural mode to the hydrodynamic mode, giving rise to the structural-hydrodynamic mode discussed in \S \ref{sec.frequency}.

The $C_p=0$ curves for $\Lambda = 0^\circ$ -- $25^\circ$ are summarized in figure \ref{fig.energy}(\emph{f}). It is seen that the trend of the top branch is similar to that observed in figure \ref{fig.bifurcation} for large-amplitude LCOs. The location of the junction between the top branch and the middle branch changes non-monotonically with $\Lambda$, which accounts for the non-monotonic behavior of the saddle-node point. In addition, figures \ref{fig.energy}(\emph{a}-\emph{e}) show that the maximum power transfer from the fluid also has a non-monotonic dependency on the sweep angle (see the shade variation of the positive $C_p$ regions as a function of the sweep angle), with an optimal sweep angle at $\Lambda=10^\circ$, which might inspire future designs of higher efficiency oscillating-foil energy harvesting devices.

\subsection{Force, moment and three-dimensional flow structures}{\label{sec.moment_PIV}}

\begin{figure}
\centering
\includegraphics[width=1\textwidth]{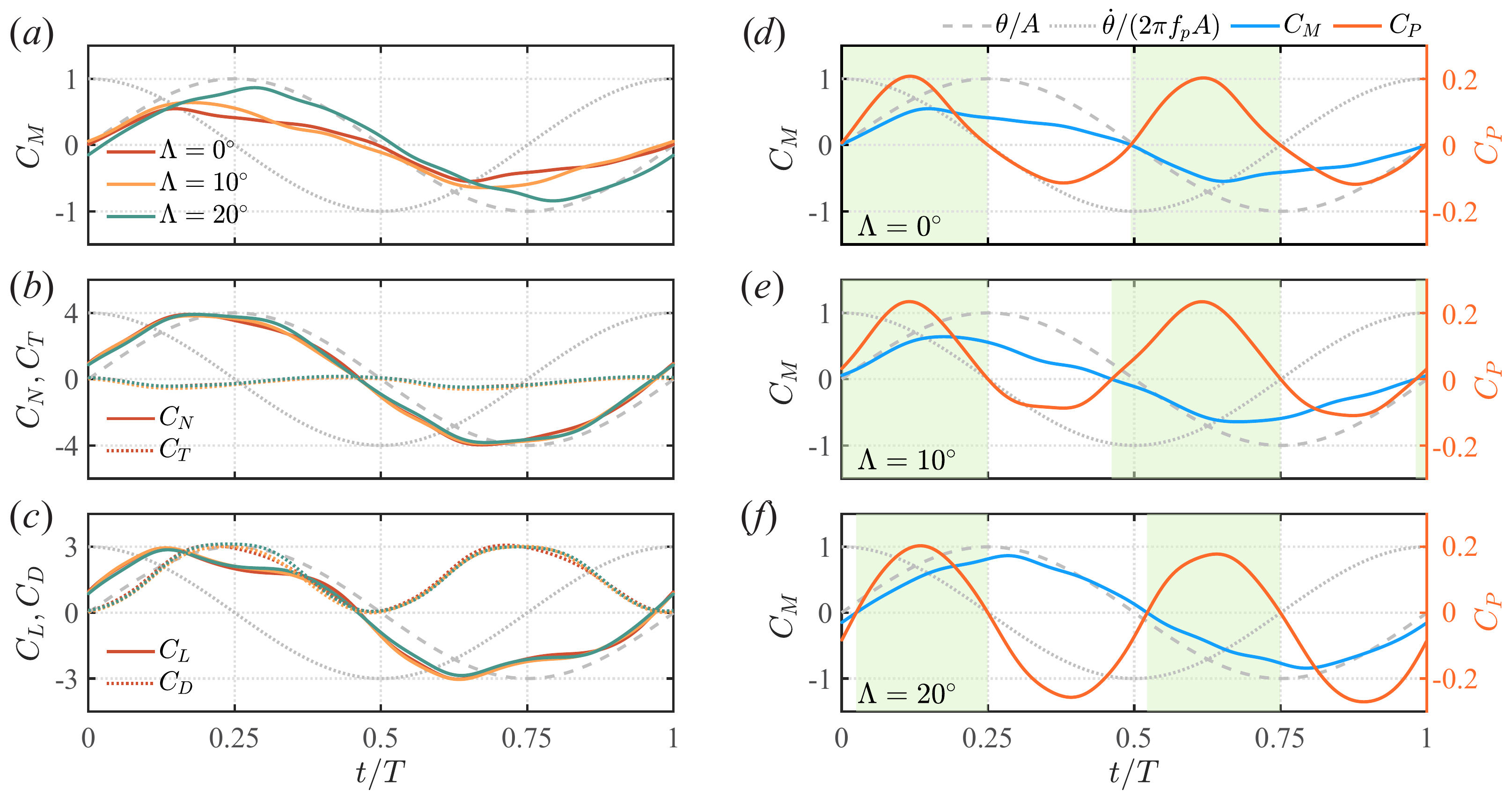}
\caption{(\emph{a}) Phase-averaged aerodynamic moment coefficients, $C_M$, and (\emph{b,c}) force coefficients, $C_N$, $C_T$, $C_L$ and $C_D$, measured at $f_p^*=0.085$, $A=1.05$ ($60^\circ$) for the $\Lambda=0^\circ$, $10^\circ$ and $20^\circ$ wings, corresponding to the black star case in figure \ref{fig.energy}(\emph{f}). (\emph{d-f}) Phase-averaged moment coefficients, $C_M$, and power coefficients, $C_P$, for $\Lambda=0^\circ$, $10^\circ$ and $20^\circ$. Green panels represent positive power input regions, where $C_P>0$. Gray dashed lines and dotted lines represent the normalized pitching angle, $\theta/A$, and pitching velocity, $\dot{\theta}/(2\pi f_p A)$, respectively. Note that $t/T=0$ corresponds to $\theta = 0$ (see the gray dashed curve).}
\label{fig.force_moment}
\end{figure}

In the previous section, \S \ref{sec.energy}, we have established the connection between prescribed oscillations and flow-induced instabilities using the energy approach. However, the question remains what causes the differences in the power coefficients measured for prescribed pitching wings with different sweep angles (figure \ref{fig.energy}). In this section, we analyze the aerodynamic force, moment and the corresponding three-dimensional flow structures to gain more insights. We focus on one pitching case, $A=1.05$ ($60^\circ$) and $f^*_p=0.085$ (i.e. the black star on figure \ref{fig.energy}\emph{f}), and three sweep angles, $\Lambda=0^\circ$, $10^\circ$ and $20^\circ$. This particular pitching kinematic is selected because it sits right on the $C_p=0$ curve for $\Lambda=0^\circ$ but in the positive $C_p$ region for $\Lambda=10^\circ$ and in the negative $C_p$ region for $\Lambda=20^\circ$ (see figure \ref{fig.energy}\emph{a,b,d,f}).

Phase-averaged coefficients of the aerodynamic moment, $C_M$, the normal force, $C_N$, the tangential force, $C_T$, the lift force, $C_L$, and the drag force, $C_D$, are plotted in figure \ref{fig.force_moment}(\emph{a-c}), respectively. Similar to the three-dimensional velocity fields, the moment and force measurements are phase-averaged over 25 cycles. We see that the moment coefficient (figure \ref{fig.force_moment}\emph{a}) behaves differently for different sweep angles, whereas the shape of other force coefficients (figure \ref{fig.force_moment}\emph{b,c}) does not change with sweep angle, resembling the trend observed in the static measurements (figure \ref{fig.static}). The observation that the wing sweep ($\Lambda = 0^\circ$ to $25^\circ$) has minimal effects on the aerodynamic force generation is non-intuitive, as one would assume that the sweep-induced spanwise flow can enhance spanwise vorticity transport in the leading-edge vortex and thereby alter the LEV stability as well as the resultant aerodynamic load. However, our measurements show the opposite, a result which is backed up by the experiments of heaving (plunging) swept wings by \citet{beem2012stabilization} ($\Lambda = 0^\circ$ to $45^\circ$) and \citet{wong2013investigation} ($\Lambda = 0^\circ$ and $\pm 45^\circ$), simulations of pitching swept wings by \citet{visbal2019effect} ($\Lambda = 0^\circ$ to $30^\circ$), and simulations of fin-like pitch-heave swept wings by \citet{zurman2021fin} ($\Lambda = 0^\circ$ to $40^\circ$), where the spanwise flow has been shown to exist but to have no effect on the aerodynamic force. We also analyze aerodynamic forces for different sweep angles and other wing kinematics and observe similar results (not shown in this manuscript). The collapse of the normal force, $C_N$, at different sweep angles suggests that the wing sweep regulates the aerodynamic moment, $C_M$, by changing the moment arm, $d_M$, as $C_M=C_N d_M$. This argument will be revisited later when we discuss the leading-edge vortex and tip vortex dynamics.

Figure \ref{fig.force_moment}(\emph{a}) shows that as the sweep angle increases, the moment coefficient, $C_M$, peaks at a later time in the cycle, and has an increased maximum value. To further analyze $C_M$ and its effects on the power coefficient, $C_P$, for different wings sweeps, we compare $C_M$ and $C_P$ for $\Lambda=0^\circ$, $10^\circ$ and $20^\circ$ in figure \ref{fig.force_moment}(\emph{d-f}), respectively. Note that here we define the power coefficient as $C_P=C_M \dot{\theta}^*$, which is different from equation \ref{eqn.power_balance} in a way that this $C_P$ is time-dependent instead of cycle-averaged, and that the power dissipated by the structure, $b^*\dot{\theta}^{*2}$ is not considered (this power dissipation is small because a small $b^*$ is used in the experiments). The normalized pitching angle, $\theta/A$, and pitching velocity, $\dot{\theta}/(2\pi f_p A)$, are also plotted for reference. We see that at the beginning of the cycle ($0 \leq t/T<0.15$), $C_M(t/T)$ grows near-linearly for all three wings. Because $\dot{\theta}>0$ for the first quarter cycle, the $x$-intercept of $C_M$ determines the starting point of the positive $C_P(t/T)$ region, corresponding to the left edge of the green panels in the figures. The $C_P>0$ region starts at $t/T=0$ for the unswept wing as $C_M$ has a near-zero $y$-intercept. For the $\Lambda=10^\circ$ swept wing, because $C_M$ has a small positive $y$-intercept, the $C_P>0$ region starts even before $t/T=0$. On the contrary, the $C_P>0$ region starts after $t/T=0$ for the $\Lambda=20^\circ$ swept wing due to a small negative $y$-intercept of $C_M$. Owing to the combined effect of an increasing $C_M$ and a decreasing $\dot{\theta}$, the power coefficient peaks around $t/T=0.125$ for all the wings. The maximum $C_P$ of the $\Lambda=10^\circ$ wing is slightly higher than that of the other two wings, due to a slightly higher $C_M$.

As the pitching cycle continues, $C_M(t/T)$ peaks around $t/T=0.15$, 0.17 and 0.28 for $\Lambda=0^\circ$, $10^\circ$ and $20^\circ$, respectively. The pitch reversal occurs at $t/T=0.25$, where $\theta$ reaches its maximum and $\dot{\theta}$ switches its sign to negative. Because the pitching velocity is now negative, the green panels terminate as $C_P$ drops below zero, suggesting that $C_M$ starts to dissipate energy into the ambient fluid. However, because $C_M$ continues to grow after $t/T=0.25$ for the $\Lambda=20^\circ$ wing, it generates a much more negative $C_P$ as compared to the wings with a lower sweep angle. Figure \ref{fig.force_moment}(\emph{a}) shows that $C_M$ decreases faster for the $\Lambda=10^\circ$ wing than the unswept wing at $0.25 \leq t/T < 0.5$. This difference results in a less negative $C_P$ for the $\Lambda=10^\circ$ wing as compared to the $\Lambda=0^\circ$ wing. The faster decrease of $C_M$ for the $\Lambda=10^\circ$ wing also makes it the first to switch back to positive power generation, where $C_M$ and $\dot{\theta}$ are both negative. The same story repeats after $t/T=0.5$ due to the symmetry of the pitching cycle. In summary, we see that subtle differences in the alignment of $C_M$ and $\dot{\theta}$ can result in considerable changes of $C_P$ for different sweep angles. The start of the $C_P>0$ region is determined by the phase of $C_M$, whereas the termination of the $C_P>0$ region depends on $\dot{\theta}$. A non-monotonic duration of the $C_P>0$ region (i.e. the size of the green panels) is observed as the sweep angle increases. The cycle-averaged power coefficient, which dictates the stability of aeroelastic systems (see \S \ref{sec.energy}), is regulated by both the amplitude and phase of the aerodynamic moment. 

\begin{figure}
\centering
\includegraphics[width=0.95\textwidth]{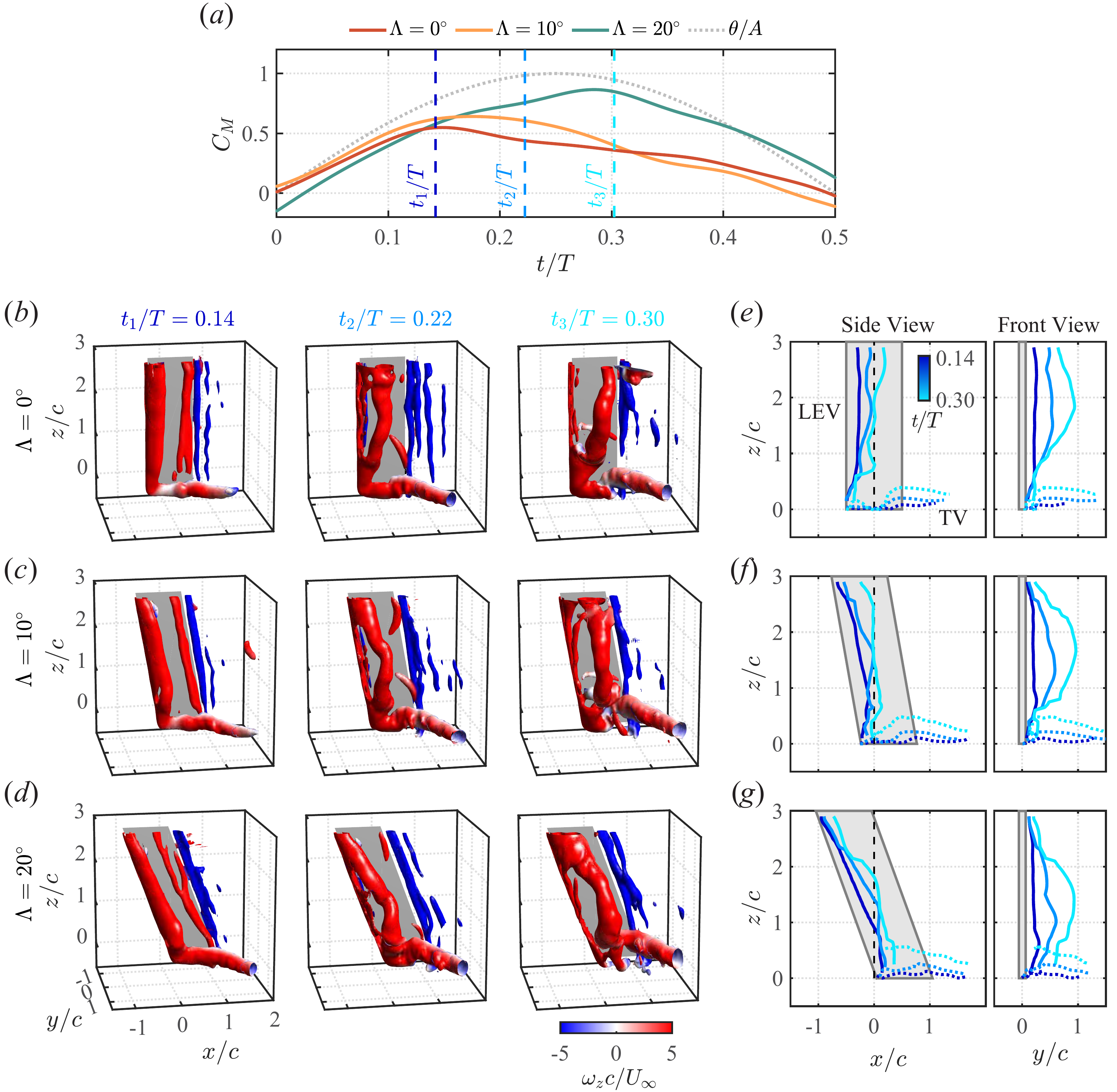}
\caption{(\emph{a}) Moment coefficients replotted from figure \ref{fig.force_moment}(\emph{a}) for half pitching cycle. Three representative time instants $t_1/T=0.14$, $t_2/T=0.22$ and $t_3/T=0.30$ are selected for studying the evolution of the leading-edge vortex (LEV) and tip vortex (TV). (\emph{b-d}) Phase-averaged three-dimensional flow structures for the $\Lambda=0^\circ$ unswept wing, and the $\Lambda=10^\circ$ and $\Lambda=20^\circ$ swept wings. The flow structures are visualized with iso-$Q$ surfaces ($Q = 50~\mathrm{s}^{-2}$) and colored by the non-dimensional spanwise vorticity, $\omega_z c/U_{\infty}$. All the flow fields are rotated by the pitching angle to keep the wing at a zero angle of attack for better visualization of the flow structures. A video capturing the three-dimensional flow structures for the entire pitching cycle can be found in the supplementary material. (\emph{e-g}) Side views and front views of the corresponding three-dimensional LEV and TV geometries. Solid curves represent LEVs and dotted lines represent TVs.}
\label{fig.PIV_traj}
\end{figure}

Next, we analyze the effect of wing sweep on the leading-edge vortex and tip vortex dynamics and the resultant impact on the aerodynamic moment. Figure \ref{fig.PIV_traj} shows (\emph{a}) the moment measurements, (\emph{b-d}) the phase-averaged three-dimensional flow structures at $t_1/T=0.14$, $t_2/T=0.22$ and $t_3/T=0.30$, and (\emph{e-g}) the corresponding leading-edge vortex and tip vortex geometries for the $\Lambda=0^\circ$, $10^\circ$ and $20^\circ$ wings. The three equally spaced time instants $t_1/T=0.14$, $t_2/T=0.22$ and $t_3/T=0.30$ are selected because they correspond to the times of the formation, growth and shedding of the leading-edge vortex. The three-dimensional flow structures are visualized using iso-$Q$ surfaces with a value of $50~\mathrm{s}^{-2}$ and colored by the non-dimensional spanwise vorticity, $\omega_z c/U_{\infty}$. In this view, the leading edge of the wing is pitching towards us, but for clarity, the flow field is always plotted with the coordinate system oriented so that the chord line is aligned with the $x-$axis. 

The initial linear growth of the moment coefficient before $t_1/T$ for all three wings corresponds to the formation of a strong leading-edge vortex, as depicted in figure \ref{fig.PIV_traj}(\emph{b-d}) at $t_1/T=0.14$, which brings the lift and moment coefficients above the static stall limit. At this stage, we see that the structure of the leading-edge vortex is similar across different wing sweeps, despite some minor variations near the wing tip. For the unswept wing, the LEV stays mostly attached along the wing span, whereas for the two swept wings, the LEV starts to detach near the tip region (see the small holes on the feeding shear layer near the wing tip). A positive vortex tube on the surface near the trailing edge is observed for all three wings, along with the negative vortex tubes shed from the trailing edge. We also observe a streamwise-oriented tip vortex wrapping over the wing tip, and this tip vortex grows stronger with the sweep angle, presumably due to the higher tip velocity associated with the larger wing sweep. Another possible cause for a stronger TV at a higher sweep angle is that the effective angle of attack becomes higher at the wing tip as the wing sweep increases.

The tracking of the vortex geometry (figure \ref{fig.PIV_traj}\emph{e-g}) provides a more quantitative measure to analyze the LEV and TV dynamics. We see that at $t_1/T=0.14$, the LEVs for all three wings are mostly aligned with the leading edge except for the tip region ($z/c=0$). For the two swept wings, the LEV also stays closer to the leading edge near the wing root ($z/c=3$). Due to the high wing sweep of the $\Lambda=20^\circ$ wing, a small portion of the LEV falls behind the pivot axis, presumably contributing to a negative moment. However, the mean distance between the LEV and the pivot axis (i.e. the LEV moment arm) stays roughly constant across different wing sweeps, potentially explaining the agreement between the $C_M$ for different wings during the linear growth region. On the other hand, the tip vortex moves downstream as the wing sweep increases due to the wing geometry. For the unswept wing and the $\Lambda=10^\circ$ swept wing, the majority of the tip vortex stays behind the pivot axis. For the $\Lambda=20^\circ$ swept wing, the TV stays entirely behind the pivot axis. As a result, the TV mostly contributes to the generation of negative moments, which counteracts the LEV moment contribution.

At $t_2/T=0.22$, figure \ref{fig.PIV_traj}(\emph{b}) and the front view of figure \ref{fig.PIV_traj}(\emph{e}) show that the LEV mostly detaches from the wing surface for the unswept wing except for a small portion near the wing tip, which stays attached. A similar flow structure was observed by \citet{yilmaz2012flow} for finite-span wings undergoing linear pitch-up motions, and by \citet{son2022leading} for high-aspect-ratio plunging wings. For the $\Lambda=10^\circ$ wing, this small portion of the attached LEV shrinks (see the front view of figure \ref{fig.PIV_traj}\emph{f}). The top portion of the LEV near the wing root is also observed to stay attached to the wing surface as compared to the $\Lambda=0^\circ$ case. For the $\Lambda=20^\circ$ wing, as shown by the front view of figure \ref{fig.PIV_traj}(\emph{g}), the attached portion of the LEV near the wing tip further shrinks and almost detaches, while the top portion of the LEV also attaches to the wing surface, similar to that observed for $\Lambda=10^\circ$. The shrinking of the LEV attached region near the wing tip as a function of the wing sweep is presumably caused by the decreased anchoring effect of the tip vortex. The shrinking of the attached LEV could also be a result of an increased effective angle of attack. The side views of figure \ref{fig.PIV_traj}(\emph{e-g}) show that the LEV moves towards the pivot axis at this time instant. The swept wing LEVs have slightly longer mean moment arms due to their attached portions near the wing root. This is more prominent for the $\Lambda=20^\circ$ wing, potentially explaining the $C_M$ of $\Lambda=20^\circ$ exceeding the other two wings at $t_2/T$. The tip vortex moves upwards and outwards with respect to the wing surface from $t_1/T$ to $t_2/T$.

During the pitch reversal ($t_3/T=0.30$), the LEV further detaches from the wing surface, and the TV also starts to detach. For the unswept wing, the LEV mostly aligns with the pivot axis except for the tip portion, which still remains attached. For the $\Lambda=10^\circ$ swept wing, the LEV also roughly aligns with the pivot axis, with both the root and the tip portions staying near the wing surface, forming a more prominent arch-like shape (see the front view of figure \ref{fig.PIV_traj}\emph{f}) as compared to the previous time step. For the $\Lambda=20^\circ$ wing, the root portion of the LEV stays attached and remains far in front of the pivot axis. The LEV detaches near the wing tip and joins with the detached TV, as shown by figure \ref{fig.PIV_traj}(\emph{d}) and the front and top views of figure \ref{fig.PIV_traj}(\emph{g}). The attachment of the LEV near the wing root and the detachment of the TV near the wing tip both contribute to a more positive $C_M$, as compared to the other two wings with lower sweep. The change of the LEV geometry as a function of the sweep angle can be associated with the arch vortices reported by \citet{visbal2019effect}. In their numerical study, it has been shown that for pitching unswept wings with free tips on both ends, an arch-type vortical structure began to form as the pitch reversal started (see their figure 6\emph{c}). In our experiments, the wings have a free tip and an endplate (i.e. a wing-body junction, or symmetry plane). Therefore, the vortical structure shown in figure \ref{fig.PIV_traj}(\emph{b}) is equivalent to one-half of the arch vortex. If we mirror the flow structures about the wing root (i.e. the endplate), we can get a complete arch vortex similar to that observed by \citet{visbal2019effect}. For swept wings, we observe one complete arch vortex for both $\Lambda=10^\circ$ (figure \ref{fig.PIV_traj}\emph{c}) and $20^\circ$ (figure \ref{fig.PIV_traj}\emph{d}). Again, if we mirror the flow structures about the wing root, there will be two arch vortices for each swept wing, agreeing well with the observation of \citet{visbal2019effect} (see their figures 10\emph{c} and 13\emph{c}). Moreover, \citet{visbal2019effect} reported that for swept wings, as $\Lambda$ increases, the vortex arch moves towards the wing tip, which is also seen in our experiments (compare the front views of figure \ref{fig.PIV_traj}\emph{e-g}).

\subsection{Insights obtained from moment partitioning}{\label{sec.FMPM}}

We have shown in the previous section, \S \ref{sec.moment_PIV}, that the aerodynamic moment is jointly determined by the leading-edge vortex and the tip vortex dynamics. Specifically, the spatial locations and geometries of the LEV and TV, as well as the vortex strength, have a combined effect on the unsteady aerodynamic moment. To obtain further insights into this complex combined effect, we use the Force and Moment Partitioning Method (FMPM) to analyze the three-dimensional flow fields. 

\begin{figure}
\centering
\includegraphics[width=0.8\textwidth]{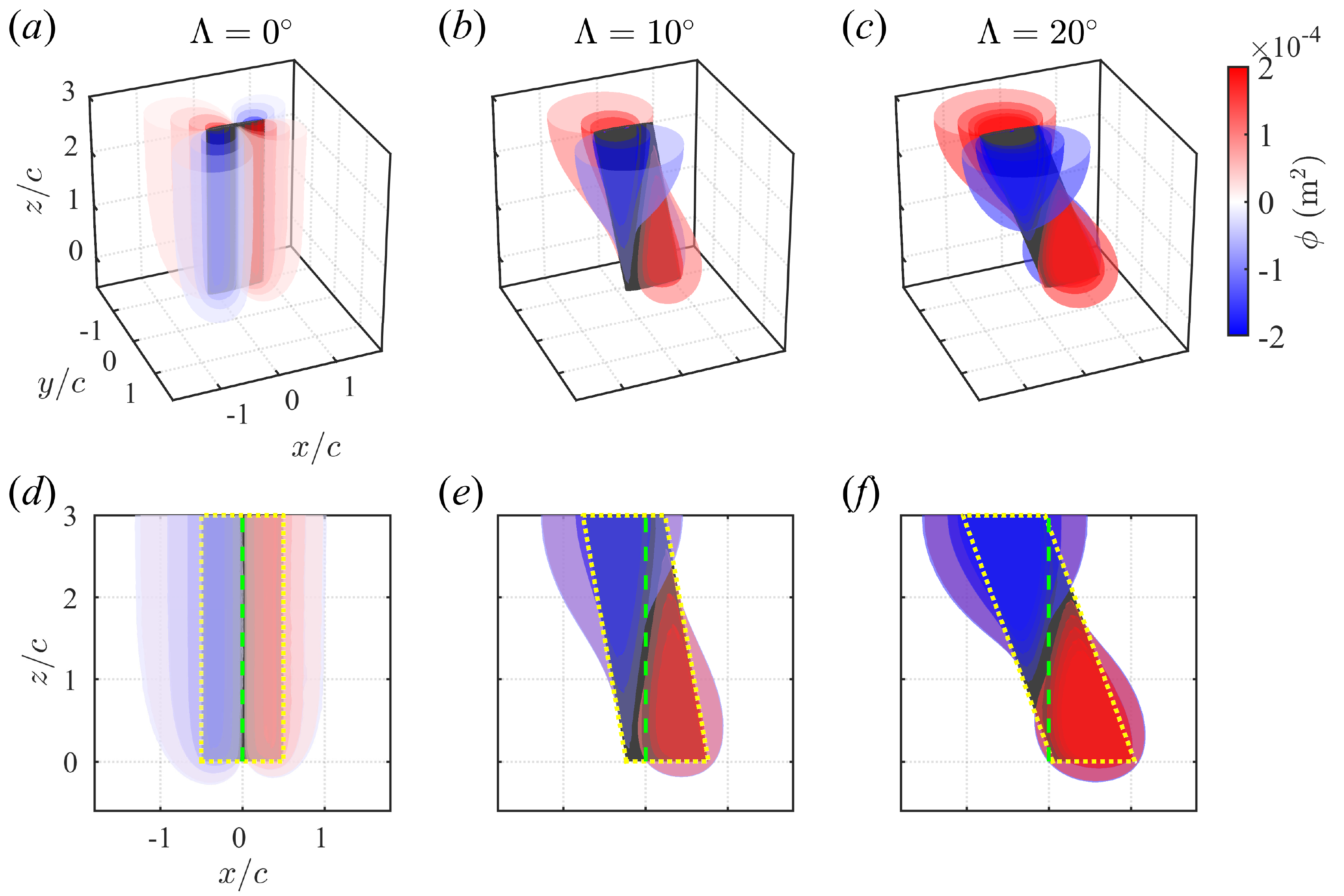}
\caption{Iso-surface plots of three-dimensional influence potentials for (\emph{a}) the $\Lambda=0^\circ$ unswept wing, (\emph{b}) the $\Lambda=10^\circ$ swept wing, and (\emph{c}) the $\Lambda=20^\circ$ swept wing. (\emph{d-f}) The corresponding side views, with the wing boundaries outlined by yellow dotted lines and the pitching axes indicated by green dashed lines.}
\label{fig.3D_phi_field}
\end{figure}

As we discussed in \S \ref{sec.FMPM_intro}, the first step of applying FMPM is to construct an `influence potential', $\phi$. We solve equation \ref{eqn.3d_phi} numerically using the MATLAB Partial Differential Equation Toolbox (Finite Element Method, code publicly available on \href{https://www.mathworks.com/matlabcentral/fileexchange/136194-force-and-moment-partitioning-influence-potential-solver-3d}{MATLAB File Exchange}). We use a 3D domain of $10c\times10c\times20c$, and a mesh resolution of $0.02c$ on the surface of the wing and $0.1c$ on the outer domain. We visualize the calculated three-dimensional influence field, $\phi$, for the $\Lambda=0^\circ$, $10^\circ$ and $20^\circ$ wings using iso-$\phi$ surfaces in figure \ref{fig.3D_phi_field}(\emph{a-c}). Figure \ref{fig.3D_phi_field}(\emph{d-f}) illustrates the corresponding side views, with the wing boundaries outlined by yellow dotted lines and the pitching axes indicated by green dashed lines. We see that for the unswept wing, the iso-$\phi$ surfaces show symmetry with respect to the pivot axis and the wing chord, resulting in a quadrant distribution of the influence field. The magnitude of $\phi$ peaks on the wing surface and decreases towards the far field. The slight asymmetry of $\phi$ with respect to the pitching axis (see figure \ref{fig.3D_phi_field}\emph{d}) is caused by the difference between the rounded leading edge and the sharp trailing edge of the NACA 0012 wing (see also the 2D influence field reported in \citet{zhu2023force}). The size of the iso-$\phi$ surfaces stays relatively constant along the wing span, except at the wing tip, where the surfaces wrap around and seal the tube. 

As the sweep angle is increased to $\Lambda=10^\circ$ and $20^\circ$, we see that the quadrant distribution of the influence field persists. However, the iso-$\phi$ surfaces form funnel-like shapes on the fore wing and teardrop shapes on the aft wing. This is caused by the variation of the effective pivot axis along the wing span. Figure \ref{fig.3D_phi_field}(\emph{e}) and (\emph{f}) show that, for swept wings, the negative $\phi$ regions extend over the entire chord near the wing root, even behind the pitching axis. Similarly, the positive $\phi$ regions (almost) cover the entire wing tip and even spill over in front of the pitching axis. As we will show next, this behavior of the $\phi$ field for swept wings will result in some non-intuitive distributions of the aerodynamic moment. In addition, the magnitude of the $\phi$ field is observed to increase with the sweep angle, due to the increase of the effective moment arm \citep{zhu2021nonlinear}.

\begin{figure}
\centering
\includegraphics[width=0.95\textwidth]{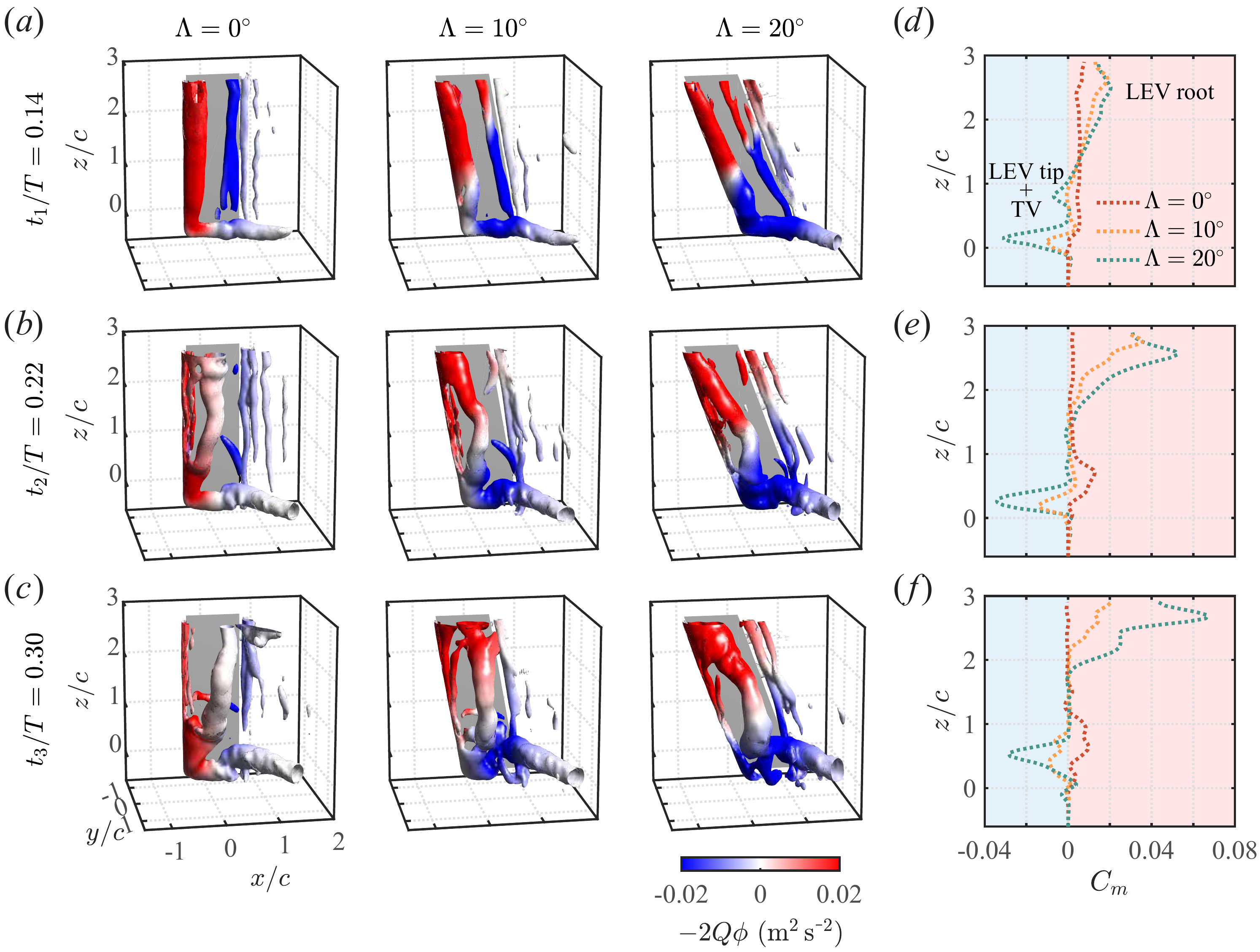}
\caption{(\emph{a-c}) Phase-averaged iso-$Q$ surfaces ($Q = 50~\mathrm{s}^{-2}$) for the $\Lambda=0^\circ$ unswept wing and the $\Lambda=10^\circ$ and $20^\circ$ swept wings, colored by the vorticity-induced moment density, $-2Q\phi$ ($\mathrm{m^2~s^{-2}}$), at $t_1/T=0.14$, $t_2/T=0.22$ and $t_3/T=0.30$. Note that the wings and flow fields are rotated in the spanwise direction to maintain a zero angle of attack, for a better view of the flow structures. (\emph{d-f}) Spanwise distributions of the vorticity-induced moment for the three wings at the three representative time instants, obtained by integrating $-2Q\phi$ at different spanwise locations.}
\label{fig.FMPM_moment}
\end{figure}

We multiply the three-dimensional $Q$ field by the influence field, $\phi$, and get the spanwise moment (density) distribution field, $-2Q\phi$. To visualize the moment distributions, we recolor the same iso-$Q$ surface plots shown in figure \ref{fig.PIV_traj} with the moment density, $-2Q\phi$, which are shown in figure \ref{fig.FMPM_moment}(\emph{a-c}). As before, the wings and flow fields are rotated by $\theta$ so that we are always looking from a viewpoint normal to the chord line, giving a better view of the flow structures. In these iso-$Q$ surface plots, red regions indicate that the vortical structure induces a positive spanwise moment, whereas blue regions represent the generation of a negative spanwise moment. In between red and blue regions, white regions have zero contribution to the spanwise moment.

At $t_1/T=0.14$ (figure \ref{fig.FMPM_moment}\emph{a}), as expected, we see that the entire LEV on the unswept wing is generating a positive moment. For the $\Lambda=10^\circ$ swept wing, however, the LEV generates a near-zero moment near the wing tip, and for the $\Lambda=20^\circ$ swept wing, the tip region of the LEV contributes a negative moment due to the non-conventional distribution of the $\phi$ field. The TV generates almost no moment for the unswept wing, but contributes a negative moment for the swept wings. The vortex tube formed near the trailing edge of the wing surface contributes entirely to negative moments for the unswept wing, but its top portion starts to generate positive moments as the sweep angle increases. The contributions of each vortical structure on the moment generation for the three wings become more clear if we plot the spanwise distribution of the vorticity-induced moment. 

By integrating the moment distribution field $-2Q\phi$ over the horizontal ($x,y$)-plane at each spanwise location, $z$, we are able to obtain the spanwise distribution of the vorticity-induced moment, shown in figure \ref{fig.FMPM_moment}(\emph{d-f}). For the unswept wing, $\Lambda=0^\circ$, figure \ref{fig.FMPM_moment}(\emph{d}) shows that the LEV generates a near-uniform positive moment across the span. As the sweep angle increases ($\Lambda=10^\circ$), the LEV generates a higher positive moment near the wing root, and the TV starts to generate a negative moment. For the $\Lambda=20^\circ$ wing, this trend persists. It is also interesting to see that the spanwise moment distribution curves for the three wings intersect around the mid span, where the effective pivot axis coincides at the mid chord. For the two swept wings, the more positive moments near the wing root counteract the negative LEV and TV contributions near the wing tip, resulting in a similar overall moment as compared to the unswept wing. The FMPM thus quantitatively explains why the three wings generate similar unsteady moments at this time instant (figure \ref{fig.PIV_traj}\emph{a}).

At $t_2/T=0.22$ (figure \ref{fig.FMPM_moment}\emph{b}), the LEV starts to detach and moves towards the pitching axis. As discussed in the previous section, \S \ref{sec.moment_PIV}, the LEV forms a half-arch for the unswept wing, with only the tip region staying attached, and a complete arch for swept wings, with both the root and tip regions staying attached. These arch-like LEV geometries, together with the special shapes of the three-dimensional influence field, lead to some special distributions of the aerodynamic moments. For the unswept wing, the color of the LEV becomes lighter as compared to the $t_1/T$ case, indicating a decreasing contribution to positive moments. However, the attached portion of the LEV still generates a positive moment as it remains attached, close to the wing, and in front of the pitching axis. Comparing the two swept wing cases, the LEV for the $\Lambda=20^\circ$ wing generates more positive moments near the wing root as compared to the $\Lambda=10^\circ$ wing due to the magnitude of the $\phi$ field (figure \ref{fig.3D_phi_field}). The TVs for the three wings behave similarly to the cases at $t_1/T$. The aft wing vortex tube on the wing surface breaks into two smaller tubes. Because of their small volumes, we do not expect them to affect the total moment generation. Figure \ref{fig.FMPM_moment}(\emph{e}) shows that the large part of the LEV does not contribute to any moment generation for the unswept wing -- only the tip region ($0\leq z/c \leq 1$) generates positive moments. As compared to $t_1/T$, the LEV generates more positive moments near the wing root for the two swept wings, especially for the $\Lambda=20^\circ$ wing, and the TV generates slightly more negative moments. The overall trend observed in figure \ref{fig.FMPM_moment}(\emph{e}) further explains the moment measurements shown in figure \ref{fig.PIV_traj}(\emph{a}), where the $\Lambda=20^\circ$ wing produces the highest $C_M$, followed by the $\Lambda=10^\circ$ wing and then the unswept wing at $t_2/T$.

At $t_3/T=0.30$ (figure \ref{fig.FMPM_moment}\emph{c}), the LEV further detaches from the wing surface. For the unswept wing, the LEV color becomes even lighter. Comparing the temporal evolution of the LEV color for the unswept wing, we see that the LEV progressively generates lower positive moments, agreeing well with the decreasing moment measurement shown in figure \ref{fig.PIV_traj}(\emph{a}). The LEV continues to generate positive moments near the root region and negative moments near the tip region for the $\Lambda=10^\circ$ swept wing, although it is largely aligned with the pivot axis (see also the side view of figure \ref{fig.PIV_traj}\emph{f}). This is again a result of the non-conventional funnel-shaped $\phi$ field near the wing root and the teardrop-like $\phi$ field near the wing tip (figure \ref{fig.3D_phi_field}\emph{b} and \emph{e}). This trend persists for the $\Lambda=20^\circ$ wing. However, the LEV generates more positive moments due to its shorter distance from the leading edge and the wing surface near the wing root. Moreover, the size of the LEV iso-$Q$ surface also becomes larger for the $\Lambda=20^\circ$ wing as compared to the previous time steps, indicating a stronger LEV and thus a higher aerodynamic moment, which explains why the $C_M$ of $\Lambda=20^\circ$ peaks around $t_3/T$ in figure \ref{fig.PIV_traj}(\emph{a}). This is also reflected in the spanwise moment plot in figure \ref{fig.FMPM_moment}(\emph{f}), where the LEV generates more positive moments for the $\Lambda=20^\circ$ wing than the $\Lambda=10^\circ$ wing. The tip vortex again behaves similarly to the previous time steps for all three wings, although it becomes less coherent and detaches from the wing surface.

It is worth noting that the integral of $-2Q\phi$ over the ($x,y$)-plane (i.e. figure \ref{fig.FMPM_moment}\emph{d-f}) also includes contributions from other vortical structures. In figure \ref{fig.FMPM_moment}(\emph{a-c}), we can see that there are four main structures on each wing: the LEV, the TV, the TEV, and the vortex tube on the aft wing surface. Figure \ref{fig.3D_phi_field} shows that the amplitude of the influence field, $\phi$, is zero near the trailing edge due to symmetry. This means that the contribution to the moment by the TEV is negligible, because $-2Q\phi$ approaches zero in this region and makes no contribution to the integrand. The aft wing vortex tube is small in size compared to the LEV and TV. In addition, it is not as coherent, because it breaks down at $t_2/T=0.22$. Therefore, we would expect its contribution to the integral to be small as well.

In summary, the Force and Moment Partitioning Method enables us to associate the complex three-dimensional vortex dynamics with the corresponding vorticity-induced moments, and quantitatively explains the mechanisms behind the observed differences in the unsteady moment generation, which further drives the pitching motion of these swept wings. These insightful analyses would not have been possible without the FMPM.

\section{Conclusion}{\label{sec.conclusion}}

In this experimental study, we have explored the nonlinear flow-induced oscillations and three-dimensional vortex dynamics of cyber-physically mounted pitching unswept and swept wings, with the pitching axis passes through the mid-chord point at the mid-span plane, and with the sweep angle varied from $0^\circ$ to $25^\circ$. At a constant flow speed, a prescribed high inertia and a small structural damping, we adjusted the wing stiffness to systematically study the onset and extinction of large-amplitude flow-induced oscillations. For the current selections of the pitching axis location and the range of the sweep angle, the amplitude response revealed subcritical Hopf bifurcations for all the unswept and swept wings, with a clustering behavior for the Hopf point and a non-monotonic saddle-node point as a function of the sweep angle. The flow-induced oscillations have been correlated with the structural oscillation mode, where the oscillations are dominated by the inertial behavior of the wing. For swept wings with high sweep angles, a hybrid oscillation mode, namely the structural-hydrodynamic mode, has been observed and characterized, in which the oscillations were regulated by both the inertial moment and the fluid moment. The onset of flow-induced oscillations (i.e. the Hopf point) has been shown to depend on the static characteristics of the wing. The non-monotonic trend of the saddle-node point against the sweep angle can be attributed to the non-monotonic power transfer between the ambient fluid and the elastic mount, which further depends on the amplitude and phase of the unsteady aerodynamic moment. Force and moment measurements have shown that, perhaps surprisingly, the wing sweep has a minimal effect on the aerodynamic forces and it was therefore inferred that the wing sweep modulates the aerodynamic moment by affecting the moment arm. Phase-averaged three-dimensional flow structures measured using stereoscopic PIV have been analyzed to characterize the dynamics of the leading-edge vortex and tip vortex. Finally, by employing the Force and Moment Partitioning Method (FMPM), we have successfully correlated the complex LEV and TV dynamics with the resultant aerodynamic moment in a quantitative manner.

In addition to reporting new observations and providing physical insights on the effects of moderate wing sweep in large-amplitude aeroelastic oscillations, the present study can serve as a source of validation data for future theoretical/computational models. Furthermore, the optimal sweep angle ($\Lambda=10^\circ$) observed for promoting flow-induced oscillations may have engineering implications. For example, one should avoid this sweep angle for aero-structure designs to stay away from aeroelastic instabilities. On the other hand, this angle could potentially be employed for developing higher-efficiency flapping-foil energy-harvesting devices. Lastly, the use of FMPM to analyze (especially three-dimensional) flow fields obtained from PIV experiments has shown great utility, and the results further demonstrated the powerful capability of this emerging method to provide valuable physical insights into vortex-dominated flows, paving the way for more applications of this method to data from future experimental and numerical studies.

\section*{Acknowledgments}
 
This work is funded by the Air Force Office of Scientific Research, Grant FA9550-21-1-0462, managed by Dr. Gregg Abate. We acknowledge the helpful discussions with Rajat Mittal, Karthik Menon, and Sushrut Kumar.

\section*{Declaration of interests}
The authors report no conflict of interest.

\bibliographystyle{jfm}
\bibliography{ref}

\begin{thebibliography}{73}
\expandafter\ifx\csname natexlab\endcsname\relax\def\natexlab#1{#1}\fi
\def\au#1{#1} \def\ed#1{#1} \def\yr#1{#1}\def\at#1{#1}\def\jt#1{\textit{#1}} \def\bt#1{#1}\def\bvol#1{\textbf{#1}} \def\vol#1{#1} \def\pg#1{#1} \def\publ#1{#1}\def\arxiv#1{#1}\def\org#1{#1}\def\st#1{\textit{#1}}

\bibitem[Beatus \& Cohen(2015)]{beatus2015wing}
{\sc \au{Beatus, T.} \& \au{Cohen, I.}} \yr{2015}  \at{Wing-pitch modulation in maneuvering fruit flies is explained by an interplay between aerodynamics and a torsional spring}.  \jt{Phys. Rev. E}  \bvol{92}~(2),  \pg{022712}.

\bibitem[Beem {\em et~al.\/}(2012)Beem, Rival \& Triantafyllou]{beem2012stabilization}
{\sc \au{Beem, H.~R.}, \au{Rival, D.~E.} \& \au{Triantafyllou, M.~S.}} \yr{2012}  \at{On the stabilization of leading-edge vortices with spanwise flow}.  \jt{Exp. Fluids}  \bvol{52}~(2),  \pg{511--517}.

\bibitem[Bergou {\em et~al.\/}(2007)Bergou, Xu \& Wang]{bergou2007passive}
{\sc \au{Bergou, A.~J.}, \au{Xu, S.} \& \au{Wang, Z.~J.}} \yr{2007}  \at{Passive wing pitch reversal in insect flight}.  \jt{J. Fluid Mech.}  \bvol{591},  \pg{321--337}.

\bibitem[Birch \& Dickinson(2001)]{birch2001spanwise}
{\sc \au{Birch, J.~M.} \& \au{Dickinson, M.~H.}} \yr{2001}  \at{Spanwise flow and the attachment of the leading-edge vortex on insect wings}.  \jt{Nature}  \bvol{412}~(6848),  \pg{729--733}.

\bibitem[Borazjani \& Daghooghi(2013)]{borazjani2013fish}
{\sc \au{Borazjani, I.} \& \au{Daghooghi, M.}} \yr{2013}  \at{The fish tail motion forms an attached leading edge vortex}.  \jt{Proc. Royal Soc. B.}  \bvol{280}~(1756),  \pg{20122071}.

\bibitem[Bottom~II {\em et~al.\/}(2016)Bottom~II, Borazjani, Blevins \& Lauder]{bottom2016hydrodynamics}
{\sc \au{Bottom~II, R.~G.}, \au{Borazjani, I.}, \au{Blevins, E.~L.} \& \au{Lauder, G.~V.}} \yr{2016}  \at{Hydrodynamics of swimming in stingrays: numerical simulations and the role of the leading-edge vortex}.  \jt{J. Fluid Mech.}  \bvol{788},  \pg{407--443}.

\bibitem[Boudreau {\em et~al.\/}(2018)Boudreau, Dumas, Rahimpour \& Oshkai]{boudreau2018experimental}
{\sc \au{Boudreau, M.}, \au{Dumas, G.}, \au{Rahimpour, M.} \& \au{Oshkai, P.}} \yr{2018}  \at{Experimental investigation of the energy extraction by a fully-passive flapping-foil hydrokinetic turbine prototype}.  \jt{J. Fluids Struct.}  \bvol{82},  \pg{446--472}.

\bibitem[Chiereghin {\em et~al.\/}(2020)Chiereghin, Bull, Cleaver \& Gursul]{chiereghin2020three}
{\sc \au{Chiereghin, N.}, \au{Bull, S.}, \au{Cleaver, D.~J.} \& \au{Gursul, I.}} \yr{2020}  \at{Three-dimensionality of leading-edge vortices on high aspect ratio plunging wings}.  \jt{Phys. Rev. Fluids}  \bvol{5}~(6),  \pg{064701}.

\bibitem[Dimitriadis \& Li(2009)]{dimitriadis2009bifurcation}
{\sc \au{Dimitriadis, G.} \& \au{Li, J.}} \yr{2009}  \at{Bifurcation behavior of airfoil undergoing stall flutter oscillations in low-speed wind tunnel}.  \jt{AIAA J.}  \bvol{47}~(11),  \pg{2577--2596}.

\bibitem[Dowell {\em et~al.\/}(1989)Dowell, Curtiss, Scanlan \& Sisto]{dowell1989modern}
{\sc \au{Dowell, E.~H.}, \au{Curtiss, H.~C.}, \au{Scanlan, R.~H.} \& \au{Sisto, F.}} \yr{1989} {\em A modern course in aeroelasticity\/}.  \publ{Springer}.

\bibitem[Eldredge \& Jones(2019)]{eldredge2019leading}
{\sc \au{Eldredge, J.~D.} \& \au{Jones, A.~R.}} \yr{2019}  \at{Leading-edge vortices: mechanics and modeling}.  \jt{Annu. Rev. Fluid Mech.}  \bvol{51},  \pg{75--104}.

\bibitem[Ellington {\em et~al.\/}(1996)Ellington, van~den Berg, Willmott \& Thomas]{ellington1996leading}
{\sc \au{Ellington, C.~P.}, \au{van~den Berg, C.}, \au{Willmott, A.~P.} \& \au{Thomas, A. L.~R.}} \yr{1996}  \at{Leading-edge vortices in insect flight}.  \jt{Nature}  \bvol{384}~(6610),  \pg{626}.

\bibitem[Gursul \& Cleaver(2019)]{gursul2019plunging}
{\sc \au{Gursul, I.} \& \au{Cleaver, D.}} \yr{2019}  \at{Plunging oscillations of airfoils and wings: Progress, opportunities, and challenges}.  \jt{AIAA J.}  \bvol{57}~(9),  \pg{3648--3665}.

\bibitem[Hartloper {\em et~al.\/}(2013)Hartloper, Kinzel \& Rival]{hartloper2013competition}
{\sc \au{Hartloper, C.}, \au{Kinzel, M.} \& \au{Rival, D.~E.}} \yr{2013}  \at{On the competition between leading-edge and tip-vortex growth for a pitching plate}.  \jt{Exp. Fluids}  \bvol{54}~(1),  \pg{1447}.

\bibitem[Hover {\em et~al.\/}(1997)Hover, Miller \& Triantafyllou]{hover1997vortex}
{\sc \au{Hover, F.~S.}, \au{Miller, S.~N.} \& \au{Triantafyllou, M.~S.}} \yr{1997}  \at{Vortex-induced vibration of marine cables: experiments using force feedback}.  \jt{J. Fluids Struct.}  \bvol{11}~(3),  \pg{307--326}.

\bibitem[Hunt {\em et~al.\/}(1988)Hunt, Wray \& Moin]{hunt1988eddies}
{\sc \au{Hunt, J. C.~R.}, \au{Wray, A.~A.} \& \au{Moin, P.}} \yr{1988}  \at{Eddies, streams, and convergence zones in turbulent flows}.  \jt{Center for Turbulence Research Report CTR-S88}  \pg{pp. 193--208}.

\bibitem[Jafferis {\em et~al.\/}(2019)Jafferis, Helbling, Karpelson \& Wood]{jafferis2019untethered}
{\sc \au{Jafferis, N.~T.}, \au{Helbling, E.~F.}, \au{Karpelson, M.} \& \au{Wood, R.~J.}} \yr{2019}  \at{Untethered flight of an insect-sized flapping-wing microscale aerial vehicle}.  \jt{Nature}  \bvol{570}~(7762),  \pg{491--495}.

\bibitem[Jeong \& Hussain(1995)]{jeong1995identification}
{\sc \au{Jeong, J.} \& \au{Hussain, F.}} \yr{1995}  \at{On the identification of a vortex}.  \jt{J. Fluid Mech.}  \bvol{285},  \pg{69--94}.

\bibitem[Jones(1947)]{jones1947effects}
{\sc \au{Jones, R.~T.}} \yr{1947}  \bt{Effects of sweepback on boundary layer and separation}. {\em Tech. Rep.\/} 1042.  \org{NACA}.

\bibitem[Kim \& Gharib(2010)]{kim2010experimental}
{\sc \au{Kim, D.} \& \au{Gharib, M.}} \yr{2010}  \at{Experimental study of three-dimensional vortex structures in translating and rotating plates}.  \jt{Exp. Fluids}  \bvol{49},  \pg{329--339}.

\bibitem[King {\em et~al.\/}(2018)King, Kumar \& Green]{king2018experimental}
{\sc \au{King, J.~T.}, \au{Kumar, R.} \& \au{Green, M.~A.}} \yr{2018}  \at{Experimental observations of the three-dimensional wake structures and dynamics generated by a rigid, bioinspired pitching panel}.  \jt{Phys. Rev. Fluids}  \bvol{3}~(3),  \pg{034701}.

\bibitem[Lentink {\em et~al.\/}(2007)Lentink, M{\"u}ller, Stamhuis, de~Kat, van Gestel, Veldhuis, Henningsson, Hedenstr{\"o}m, Videler \& van Leeuwen]{lentink2007swifts}
{\sc \au{Lentink, D.}, \au{M{\"u}ller, U.~K.}, \au{Stamhuis, E.~J.}, \au{de~Kat, R.}, \au{van Gestel, W.}, \au{Veldhuis, L. L.~M.}, \au{Henningsson, P.}, \au{Hedenstr{\"o}m, A.}, \au{Videler, J.~J.} \& \au{van Leeuwen, J.~L.}} \yr{2007}  \at{How swifts control their glide performance with morphing wings}.  \jt{Nature}  \bvol{446}~(7139),  \pg{1082--1085}.

\bibitem[Li {\em et~al.\/}(2020{\natexlab{{\em a\/}}})Li, Wang, Graham \& Zhao]{li2020vortex}
{\sc \au{Li, J.}, \au{Wang, Y.}, \au{Graham, M.} \& \au{Zhao, X.}} \yr{2020{\natexlab{{\em a\/}}}}  \at{Vortex moment map for unsteady incompressible viscous flows}.  \jt{J. Fluid Mech.}  \bvol{891},  \pg{A13}.

\bibitem[Li \& Wu(2018)]{li2018vortex}
{\sc \au{Li, J.} \& \au{Wu, Z.-N.}} \yr{2018}  \at{Vortex force map method for viscous flows of general airfoils}.  \jt{J. Fluid Mech.}  \bvol{836},  \pg{145--166}.

\bibitem[Li {\em et~al.\/}(2020{\natexlab{{\em b\/}}})Li, Zhao \& Graham]{li2020vortex3d}
{\sc \au{Li, J.}, \au{Zhao, X.} \& \au{Graham, M.}} \yr{2020{\natexlab{{\em b\/}}}}  \at{Vortex force maps for three-dimensional unsteady flows with application to a delta wing}.  \jt{J. Fluid Mech.}  \bvol{900}.

\bibitem[Long \& Nipper(1996)]{long1996importance}
{\sc \au{Long, J.~H.} \& \au{Nipper, K.~S.}} \yr{1996}  \at{The importance of body stiffness in undulatory propulsion}.  \jt{Am. Zool.}  \bvol{36}~(6),  \pg{678--694}.

\bibitem[Mackowski \& Williamson(2011)]{mackowski2011developing}
{\sc \au{Mackowski, A.~W.} \& \au{Williamson, C. H.~K.}} \yr{2011}  \at{Developing a cyber-physical fluid dynamics facility for fluid-structure interaction studies}.  \jt{J. Fluids Struct.}  \bvol{27}~(5-6),  \pg{748--757}.

\bibitem[McCroskey(1982)]{mccroskey1982unsteady}
{\sc \au{McCroskey, W.~J.}} \yr{1982}  \at{Unsteady airfoils}.  \jt{Annu. Rev. Fluid Mech.}  \bvol{14}~(1),  \pg{285--311}.

\bibitem[Menon {\em et~al.\/}(2022)Menon, Kumar \& Mittal]{menon2022contribution}
{\sc \au{Menon, K.}, \au{Kumar, S.} \& \au{Mittal, R.}} \yr{2022}  \at{Contribution of spanwise and cross-span vortices to the lift generation of low-aspect-ratio wings: Insights from force partitioning}.  \jt{Phys. Rev. Fluids}  \bvol{7}~(11),  \pg{114102}.

\bibitem[Menon \& Mittal(2019)]{menon2019flow}
{\sc \au{Menon, K.} \& \au{Mittal, R.}} \yr{2019}  \at{Flow physics and dynamics of flow-induced pitch oscillations of an airfoil}.  \jt{J. Fluid Mech.}  \bvol{877},  \pg{582--613}.

\bibitem[Menon \& Mittal(2021{\natexlab{{\em a\/}}})]{menon2021initiation}
{\sc \au{Menon, K.} \& \au{Mittal, R.}} \yr{2021{\natexlab{{\em a\/}}}}  \at{On the initiation and sustenance of flow-induced vibration of cylinders: insights from force partitioning}.  \jt{J. Fluid Mech.}  \bvol{907}.

\bibitem[Menon \& Mittal(2021{\natexlab{{\em b\/}}})]{menon2021quantitative}
{\sc \au{Menon, K.} \& \au{Mittal, R.}} \yr{2021{\natexlab{{\em b\/}}}}  \at{Quantitative analysis of the kinematics and induced aerodynamic loading of individual vortices in vortex-dominated flows: a computation and data-driven approach}.  \jt{J. Comput. Phys.}  \bvol{443},  \pg{110515}.

\bibitem[Menon \& Mittal(2021{\natexlab{{\em c\/}}})]{menon2021significance}
{\sc \au{Menon, K.} \& \au{Mittal, R.}} \yr{2021{\natexlab{{\em c\/}}}}  \at{Significance of the strain-dominated region around a vortex on induced aerodynamic loads}.  \jt{J. Fluid Mech.}  \bvol{918},  \pg{R3}.

\bibitem[Moriche {\em et~al.\/}(2017)Moriche, Flores \& Garc{\'\i}a-Villalba]{moriche2017aerodynamic}
{\sc \au{Moriche, M.}, \au{Flores, O.} \& \au{Garc{\'\i}a-Villalba, M.}} \yr{2017}  \at{On the aerodynamic forces on heaving and pitching airfoils at low reynolds number}.  \jt{J. Fluid Mech.}  \bvol{828},  \pg{395--423}.

\bibitem[Morse \& Williamson(2009)]{morse2009prediction}
{\sc \au{Morse, T.~L.} \& \au{Williamson, C. H.~K.}} \yr{2009}  \at{Prediction of vortex-induced vibration response by employing controlled motion}.  \jt{J. Fluid Mech.}  \bvol{634},  \pg{5--39}.

\bibitem[Mulleners \& Raffel(2012)]{mulleners2012onset}
{\sc \au{Mulleners, K.} \& \au{Raffel, M.}} \yr{2012}  \at{The onset of dynamic stall revisited}.  \jt{Exp. Fluids}  \bvol{52}~(3),  \pg{779--793}.

\bibitem[Negi {\em et~al.\/}(2021)Negi, Hanifi \& Henningson]{negi2021onset}
{\sc \au{Negi, P.~S.}, \au{Hanifi, A.} \& \au{Henningson, D.~S.}} \yr{2021}  \at{On the onset of aeroelastic pitch-oscillations of a naca0012 wing at transitional reynolds numbers}.  \jt{J. Fluids Struct.}  \bvol{105},  \pg{103344}.

\bibitem[Onoue \& Breuer(2016)]{onoue2016vortex}
{\sc \au{Onoue, K.} \& \au{Breuer, K.~S.}} \yr{2016}  \at{Vortex formation and shedding from a cyber-physical pitching plate}.  \jt{J. Fluid Mech.}  \bvol{793},  \pg{229--247}.

\bibitem[Onoue \& Breuer(2017)]{onoue2017scaling}
{\sc \au{Onoue, K.} \& \au{Breuer, K.~S.}} \yr{2017}  \at{A scaling for vortex formation on swept and unswept pitching wings}.  \jt{J. Fluid Mech.}  \bvol{832},  \pg{697--720}.

\bibitem[Onoue {\em et~al.\/}(2015)Onoue, Song, Strom \& Breuer]{onoue2015large}
{\sc \au{Onoue, K.}, \au{Song, A.}, \au{Strom, B.} \& \au{Breuer, K.~S.}} \yr{2015}  \at{Large amplitude flow-induced oscillations and energy harvesting using a cyber-physical pitching plate}.  \jt{J. Fluids Struct.}  \bvol{55},  \pg{262--275}.

\bibitem[Polhamus(1971)]{polhamus1971predictions}
{\sc \au{Polhamus, E.~C.}} \yr{1971}  \at{Predictions of vortex-lift characteristics by a leading-edge suction analogy}.  \jt{J. Aircr.}  \bvol{8}~(4),  \pg{193--199}.

\bibitem[Quartapelle \& Napolitano(1983)]{quartapelle1983force}
{\sc \au{Quartapelle, L.} \& \au{Napolitano, M.}} \yr{1983}  \at{Force and moment in incompressible flows}.  \jt{AIAA J.}  \bvol{21}~(6),  \pg{911--913}.

\bibitem[Quinn \& Lauder(2021)]{quinn2021tunable}
{\sc \au{Quinn, D.} \& \au{Lauder, G.}} \yr{2021}  \at{Tunable stiffness in fish robotics: mechanisms and advantages}.  \jt{Bioinspir. Biomim.}  \bvol{17}~(1),  \pg{011002}.

\bibitem[Rao(1995)]{rao1995mechanical}
{\sc \au{Rao, S.~S.}} \yr{1995} {\em Mechanical Vibrations\/}.  \publ{Addison-Wesley}.

\bibitem[Ribeiro {\em et~al.\/}(2022)Ribeiro, Yeh, Zhang \& Taira]{ribeiro2022wing}
{\sc \au{Ribeiro, J. H.~M.}, \au{Yeh, C.-A.}, \au{Zhang, K.} \& \au{Taira, K.}} \yr{2022}  \at{Wing sweep effects on laminar separated flows}.  \jt{J. Fluid Mech.}  \bvol{950},  \pg{A23}.

\bibitem[Rival {\em et~al.\/}(2014)Rival, Kriegseis, Schaub, Widmann \& Tropea]{rival2014characteristic}
{\sc \au{Rival, D.~E.}, \au{Kriegseis, J.}, \au{Schaub, P.}, \au{Widmann, A.} \& \au{Tropea, C.}} \yr{2014}  \at{Characteristic length scales for vortex detachment on plunging profiles with varying leading-edge geometry}.  \jt{Exp. Fluids}  \bvol{55}~(1),  \pg{1--8}.

\bibitem[Shyy {\em et~al.\/}(2010)Shyy, Aono, Chimakurthi, Trizila, Kang, Cesnik \& Liu]{shyy2010recent}
{\sc \au{Shyy, W.}, \au{Aono, H.}, \au{Chimakurthi, S.~K.}, \au{Trizila, P.}, \au{Kang, C.-K.}, \au{Cesnik, C. E.~S.} \& \au{Liu, H.}} \yr{2010}  \at{Recent progress in flapping wing aerodynamics and aeroelasticity}.  \jt{Prog. Aerosp. Sci.}  \bvol{46}~(7),  \pg{284--327}.

\bibitem[Son {\em et~al.\/}(2022{\natexlab{{\em a\/}}})Son, Gao, Gursul, Cantwell, Wang \& Sherwin]{son2022leading}
{\sc \au{Son, O.}, \au{Gao, A.-K.}, \au{Gursul, I.}, \au{Cantwell, C.~D.}, \au{Wang, Z.} \& \au{Sherwin, S.~J.}} \yr{2022{\natexlab{{\em a\/}}}}  \at{Leading-edge vortex dynamics on plunging airfoils and wings}.  \jt{J. Fluid Mech.}  \bvol{940},  \pg{A28}.

\bibitem[Son {\em et~al.\/}(2022{\natexlab{{\em b\/}}})Son, Wang \& Gursul]{son2022dynamics}
{\sc \au{Son, O.}, \au{Wang, Z.} \& \au{Gursul, I.}} \yr{2022{\natexlab{{\em b\/}}}}  \at{Dynamics of tip vortices on plunging wings}.  \jt{Aerosp. Sci. Technol.}  \bvol{128},  \pg{107761}.

\bibitem[Strogatz(1994)]{strogatz1994nonlinear}
{\sc \au{Strogatz, S.~H.}} \yr{1994} {\em Nonlinear Dynamics and Chaos: With Applications to Physics, Biology, Chemistry, and Engineering\/}.  \publ{Perseus Books}.

\bibitem[Su \& Breuer(2019)]{su2019resonant}
{\sc \au{Su, Y} \& \au{Breuer, K.~S.}} \yr{2019}  \at{Resonant response and optimal energy harvesting of an elastically mounted pitching and heaving hydrofoil}.  \jt{Phys. Rev. Fluids}  \bvol{4}~(6),  \pg{064701}.

\bibitem[Taira \& Colonius(2009)]{taira2009three}
{\sc \au{Taira, K.} \& \au{Colonius, T.}} \yr{2009}  \at{Three-dimensional flows around low-aspect-ratio flat-plate wings at low reynolds numbers}.  \jt{J. Fluid Mech.}  \bvol{623},  \pg{187--207}.

\bibitem[Tong {\em et~al.\/}(2022)Tong, Wu, Chen, Wang, Du, Tan \& Yu]{tong2022design}
{\sc \au{Tong, R.}, \au{Wu, Z.}, \au{Chen, D.}, \au{Wang, J.}, \au{Du, S.}, \au{Tan, M.} \& \au{Yu, J.}} \yr{2022}  \at{Design and optimization of an untethered high-performance robotic tuna}.  \jt{IEEE/ASME Trans. Mechatron.}  \bvol{27}~(5),  \pg{4132--4142}.

\bibitem[Visbal \& Garmann(2019)]{visbal2019effect}
{\sc \au{Visbal, M.~R.} \& \au{Garmann, D.~J.}} \yr{2019}  \at{Effect of sweep on dynamic stall of a pitching finite-aspect-ratio wing}.  \jt{AIAA J.}  \bvol{57}~(8),  \pg{3274--3289}.

\bibitem[Wang(2005)]{wang2005dissecting}
{\sc \au{Wang, Z.~J.}} \yr{2005}  \at{Dissecting insect flight}.  \jt{Annu. Rev. Fluid Mech.}  \bvol{37},  \pg{183--210}.

\bibitem[Wojcik \& Buchholz(2014)]{wojcik2014vorticity}
{\sc \au{Wojcik, C.~J.} \& \au{Buchholz, J. H.~J.}} \yr{2014}  \at{Vorticity transport in the leading-edge vortex on a rotating blade}.  \jt{J. Fluid Mech.}  \bvol{743},  \pg{249}.

\bibitem[Wong {\em et~al.\/}(2013)Wong, Kriegseis \& Rival]{wong2013investigation}
{\sc \au{Wong, J.~G.}, \au{Kriegseis, J.} \& \au{Rival, D.~E.}} \yr{2013}  \at{An investigation into vortex growth and stabilization for two-dimensional plunging and flapping plates with varying sweep}.  \jt{J. Fluids Struct.}  \bvol{43},  \pg{231--243}.

\bibitem[Wong \& Rival(2015)]{wong2015determining}
{\sc \au{Wong, J.~G.} \& \au{Rival, D.~E.}} \yr{2015}  \at{Determining the relative stability of leading-edge vortices on nominally two-dimensional flapping profiles}.  \jt{J. Fluid Mech.}  \bvol{766},  \pg{611}.

\bibitem[Wu {\em et~al.\/}(2019)Wu, Nowak \& Breuer]{wu2019scaling}
{\sc \au{Wu, K.~S.}, \au{Nowak, J.} \& \au{Breuer, K.~S.}} \yr{2019}  \at{Scaling of the performance of insect-inspired passive-pitching flapping wings}.  \jt{J. R. Soc. Interface}  \bvol{16}~(161),  \pg{20190609}.

\bibitem[Xiao \& Zhu(2014)]{xiao2014review}
{\sc \au{Xiao, Q.} \& \au{Zhu, Q.}} \yr{2014}  \at{A review on flow energy harvesters based on flapping foils}.  \jt{J. Fluids Struct.}  \bvol{46},  \pg{174--191}.

\bibitem[Yilmaz \& Rockwell(2012)]{yilmaz2012flow}
{\sc \au{Yilmaz, T.~O.} \& \au{Rockwell, D.}} \yr{2012}  \at{Flow structure on finite-span wings due to pitch-up motion}.  \jt{J. Fluid Mech.}  \bvol{691},  \pg{518}.

\bibitem[Young {\em et~al.\/}(2014)Young, Lai \& Platzer]{young2014review}
{\sc \au{Young, J.}, \au{Lai, J. C.~S.} \& \au{Platzer, M.~F.}} \yr{2014}  \at{A review of progress and challenges in flapping foil power generation}.  \jt{Prog. Aerosp. Sci.}  \bvol{67},  \pg{2--28}.

\bibitem[Yuan {\em et~al.\/}(2015)Yuan, Poirel, Wang \& Benaissa]{yuan2015effect}
{\sc \au{Yuan, W.}, \au{Poirel, D.}, \au{Wang, B.} \& \au{Benaissa, A.}} \yr{2015}  \at{Effect of freestream turbulence on airfoil limit-cycle oscillations at transitional reynolds numbers}.  \jt{J. Aircr.}  \bvol{52}~(4),  \pg{1214--1225}.

\bibitem[Zhang {\em et~al.\/}(2015)Zhang, Hedrick \& Mittal]{zhang2015centripetal}
{\sc \au{Zhang, C.}, \au{Hedrick, T.~L.} \& \au{Mittal, R.}} \yr{2015}  \at{Centripetal acceleration reaction: an effective and robust mechanism for flapping flight in insects}.  \jt{PloS One}  \bvol{10}~(8),  \pg{e0132093}.

\bibitem[Zhang {\em et~al.\/}(2020{\natexlab{{\em a\/}}})Zhang, Hayostek, Amitay, Burtsev, Theofilis \& Taira]{zhang2020laminar}
{\sc \au{Zhang, K.}, \au{Hayostek, S.}, \au{Amitay, M.}, \au{Burtsev, A.}, \au{Theofilis, V.} \& \au{Taira, K.}} \yr{2020{\natexlab{{\em a\/}}}}  \at{Laminar separated flows over finite-aspect-ratio swept wings}.  \jt{J. Fluid Mech.}  \bvol{905},  \pg{R1}.

\bibitem[Zhang {\em et~al.\/}(2020{\natexlab{{\em b\/}}})Zhang, Hayostek, Amitay, He, Theofilis \& Taira]{zhang2020formation}
{\sc \au{Zhang, K.}, \au{Hayostek, S.}, \au{Amitay, M.}, \au{He, W.}, \au{Theofilis, V.} \& \au{Taira, K.}} \yr{2020{\natexlab{{\em b\/}}}}  \at{On the formation of three-dimensional separated flows over wings under tip effects}.  \jt{J. Fluid Mech.}  \bvol{895},  \pg{A9}.

\bibitem[Zhang \& Taira(2022)]{zhang2022laminar}
{\sc \au{Zhang, K.} \& \au{Taira, K.}} \yr{2022}  \at{Laminar vortex dynamics around forward-swept wings}.  \jt{Phys. Rev. Fluids}  \bvol{7}~(2),  \pg{024704}.

\bibitem[Zhong {\em et~al.\/}(2021{\natexlab{{\em a\/}}})Zhong, Han, Moored \& Quinn]{zhong2021aspect}
{\sc \au{Zhong, Q.}, \au{Han, T.}, \au{Moored, K.~W.} \& \au{Quinn, D.~B.}} \yr{2021{\natexlab{{\em a\/}}}}  \at{Aspect ratio affects the equilibrium altitude of near-ground swimmers}.  \jt{J. Fluid Mech.}  \bvol{917}.

\bibitem[Zhong {\em et~al.\/}(2021{\natexlab{{\em b\/}}})Zhong, Zhu, Fish, Kerr, Downs, Bart-Smith \& Quinn]{zhong2021tunable}
{\sc \au{Zhong, Q.}, \au{Zhu, J.}, \au{Fish, F.~E.}, \au{Kerr, S.~J.}, \au{Downs, A.~M.}, \au{Bart-Smith, H.} \& \au{Quinn, D.~B.}} \yr{2021{\natexlab{{\em b\/}}}}  \at{Tunable stiffness enables fast and efficient swimming in fish-like robots}.  \jt{Sci. Robot.}  \bvol{6}~(57),  \pg{eabe4088}.

\bibitem[Zhu {\em et~al.\/}(2023)Zhu, Lee, Kumar, Menon, Mittal \& Breuer]{zhu2023force}
{\sc \au{Zhu, Y.}, \au{Lee, H.}, \au{Kumar, S.}, \au{Menon, K.}, \au{Mittal, R.} \& \au{Breuer, K.}} \yr{2023}  \at{Force moment partitioning and scaling analysis of vortices shed by a 2d pitching wing in quiescent fluid}.  \jt{Exp. Fluids}  \bvol{64}~(10),  \pg{158}.

\bibitem[Zhu {\em et~al.\/}(2021)Zhu, Mathai \& Breuer]{zhu2021nonlinear}
{\sc \au{Zhu, Y.}, \au{Mathai, V.} \& \au{Breuer, K.}} \yr{2021}  \at{Nonlinear fluid damping of elastically mounted pitching wings in quiescent water}.  \jt{J. Fluid Mech.}  \bvol{923},  \pg{R2}.

\bibitem[Zhu {\em et~al.\/}(2020)Zhu, Su \& Breuer]{zhu2020nonlinear}
{\sc \au{Zhu, Y.}, \au{Su, Y.} \& \au{Breuer, K.}} \yr{2020}  \at{Nonlinear flow-induced instability of an elastically mounted pitching wing}.  \jt{J. Fluid Mech.}  \bvol{899},  \pg{A35}.

\bibitem[Zurman-Nasution {\em et~al.\/}(2021)Zurman-Nasution, Ganapathisubramani \& Weymouth]{zurman2021fin}
{\sc \au{Zurman-Nasution, A.~N.}, \au{Ganapathisubramani, B.} \& \au{Weymouth, G.~D.}} \yr{2021}  \at{Fin sweep angle does not determine flapping propulsive performance}.  \jt{J. R. Soc. Interface}  \bvol{18}~(178),  \pg{20210174}.

\end{thebibliography}

\end{document}